\documentclass[
reprint, 
aps, 
amsmath, 
amssymb,
superscriptaddress
]{revtex4-1}

\usepackage[caption=false]{subfig}

\usepackage{lmodern} 
\usepackage[bookmarks, colorlinks=false]{hyperref}
\usepackage{mathtools}
\hypersetup{colorlinks=true, allcolors=blue}

\usepackage{multirow}

\usepackage{graphicx,tikz}
\usepackage{dcolumn}
\usepackage{bm}
\usepackage{siunitx} 

\renewcommand{\vec}[1]{\boldsymbol{#1}}

\def\hgeom#1{{}_{2}F_{1}\!\left(\textstyle{#1}\right)}

\def\kB{k_\mathrm{B}}

\def\Ang{\mbox{\AA}}

\def\dint#1{\text{d}#1}

\begin{document}

\title{Statistical mechanics of kinks on a gliding screw dislocation}

\author{Max Boleininger}%
\email{max.boleininger@ukaea.uk}%
\affiliation{CCFE, Culham Science Centre, UK Atomic Energy Authority, Abingdon, Oxfordshire OX14 3DB, United Kingdom
}%
\author{Martin Gallauer}%
\email{gallauer@maths.ox.ac.uk}%
\affiliation{Mathematical Institute, University of Oxford, Oxford OX2 6GG, United Kingdom
}%
\author{Sergei L. Dudarev}%
\email{sergei.dudarev@ukaea.uk}%
\affiliation{CCFE, Culham Science Centre, UK Atomic Energy Authority, Abingdon, Oxfordshire OX14 3DB, United Kingdom
}%
\author{Thomas D. Swinburne}%
\email{swinburne@cinam.univ-mrs.fr}%
\affiliation{Aix-Marseille Universit\'{e}, CNRS, CINaM UMR 7325, Campus de Luminy, 13288 Marseille, France}%
\author{Daniel R. Mason}%
\email{daniel.mason@ukaea.uk}%
\affiliation{CCFE, Culham Science Centre, UK Atomic Energy Authority, Abingdon, Oxfordshire OX14 3DB, United Kingdom
}%
\author{Danny Perez}%
\email{danny\_perez@lanl.gov}%
\affiliation{Theoretical Division T-1, Los Alamos National Laboratory, Los Alamos, NM 87545, USA
}%
\date{\today}

\begin{abstract}
The ability of a body-centered cubic metal to deform plastically is limited by the thermally activated glide motion of screw dislocations, which are line defects with a mobility exhibiting complex dependence on temperature, stress, and dislocation segment length. We derive an analytical expression for the velocity of dislocation glide, based on a statistical mechanics argument, and identify an apparent phase transition marked by a critical temperature above which the activation energy for glide effectively halves, changing from the formation energy of a double kink to that of a single kink. The analysis is in quantitative agreement with direct kinetic Monte Carlo simulations.
\end{abstract}

\maketitle

The rate of plastic deformation of a body-centred-cubic (bcc) metal depends strongly on temperature, with pure iron notoriously becoming brittle below freezing \cite{tanaka2008brittle}. The temperature of this brittle-to-ductile transition, where the metal is brittle at a low temperature and is ductile at an elevated temperature, is raised by up to hundreds of degrees after irradiation by highly energetic particles, which leads to stringent requirements on the minimum operating temperature of ferritic steels and other bcc materials for technological applications in a radiation environment \cite{zinkle2012nuclear, federici2017european,  pintsuk2019european}. In order to predict how the mechanical properties vary over the service lifetime of a structural component exposed to radiation at elevated temperatures, it is desirable to develop an explicit model, relating microstructural changes to plastic deformation.

The principal rate-limiting mechanism for plastic deformation in bcc metals is the thermally activated motion (glide) of screw dislocations \cite{brunner2000comparison}, which are topological line defects in the crystal lattice, acting as carriers of plastic deformation. A screw dislocation can be approximated by an elastic line placed in a periodic Peierls potential \cite{peierls1940size, nabarro1947dislocations}, driven by shear stress and by random thermal noise from the surrounding crystal lattice \cite{braun1998nonlinear, fitzgerald2016kink}. At low stresses, screw dislocations move by the kink mechanism, in which thermal fluctuations first lead to the nucleation of a pair of kinks into the next valley of the Peierls potential, which then subsequently diffuse athermally along the dislocation \cite{swinburne2013theory} to advance the rest of the line by a discrete amount. As a result, one would expect the activation energy characterising the plastic strain rate to be close to the formation free energy of a kink pair $2f_k$. However, in an apparent contradiction to this argument, experimental observations of the brittle-to-ductile transition in high-purity bcc metals appear to exhibit activation energies closer to the formation free energy of a single kink $f_k$ \cite{giannattasio2007empirical, tanaka2008brittle, abernethy2019effects, swinburne2018kink}.

\begin{figure}[t]
\includegraphics[width=.45\columnwidth]{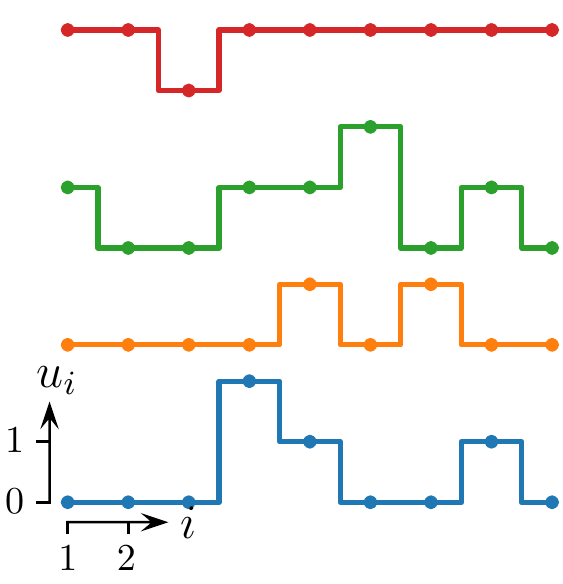}
\hspace{.25cm}
\includegraphics[width=.48\columnwidth]{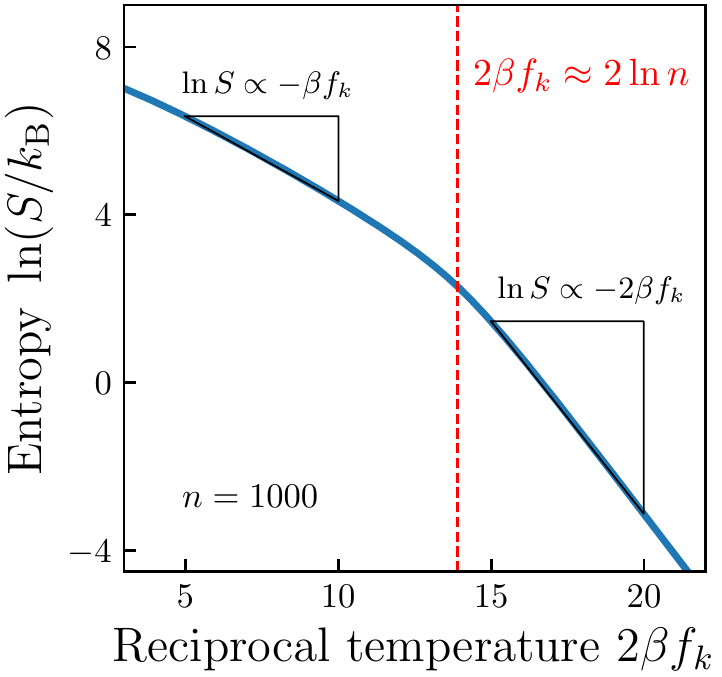} \\ 
\caption{Left: Example microstates of a screw dislocation ($n=9$). Right: Entropy as a function of reciprocal temperature $\beta = (k_B T)^{-1}$ computed using the free energy model \eqref{eq:freeenergy}. The entropy becomes extrinsic for temperatures above a certain critical value, indicated by a dashed line in the right panel, see equation \eqref{eq:ntransition}.}
\label{fig:figone} 
\end{figure}

Swinburne and Dudarev \cite{swinburne2018kink} offered a resolution to this contradiction by proposing a \textit{critical transition} in the activation energy for dislocation glide. By analyzing kink populations on a dislocation in thermal equilibrium, they concluded that a dislocation segment longer than a certain temperature-dependent critical length $L^*$ glides with an activation energy of $f_k$, while shorter dislocation segments glide with an activation energy of $2 f_k$. This conclusion is also reached in the kink diffusion model by Hirth and Lothe \cite{hirth1982theory}, based on an argument that kinks start annihilating if the line length exceeds the mean spacing between thermal kinks $\sim L^*$ \cite{maeda1993dislocation,Robertson2011}. The segment length-dependent mobility affects the ability of a screw dislocation to unpin from obstacles \cite{swinburne2018kink}, and as such has major implications for the predictive interpretation of experiments on obstacle hardening and embrittlement. Yet despite the extensive use of screw dislocation mobility laws in coarse-grained methods for modelling plastic deformation \cite{cai2004mobility, monnet2013dislocation, vattre2014modelling, cereceda2016unraveling}, to our knowledge there is still no suitable analytical expression able to capture the full complexity of dislocation mobility, consistent with the microscopic statistical mechanics description of a fluctuating dislocation line, including the critical transition noted above.

In this letter, we give a closed-form expression for the glide velocity of a screw dislocation in the kink-limited regime as a function of temperature, resolved shear stress, and segment length. This analytical model explicitly captures the critical transition, and is in quantitative and qualitative agreement with kinetic Monte Carlo (kMC) dislocation dynamics simulations.

The model describes a screw dislocation, moving in a glide plane, by a set of $n \in \mathbb{Z}^+$
displacement variables $u_i$ adopting integer values and satisfying the Born-{von Karman} periodic boundary condition $u_{n+1} = u_1$. The dislocation line is spanned by points $(i b, h u_i)$, where $i$ is an integer, $b$ is the Burgers vector length and $h$ is the distance between the adjacent Peierls valleys. 
A site $i$ may accommodate any number of left kinks $(u_{i+1} - u_i > 0)$ or right kinks $(u_{i+1} - u_i < 0)$, or no kinks at all $(u_{i+1} - u_i = 0)$. The positioning of kinks along the line is enough to uniquely characterise a line configuration, and in what follows we define the set of microstates as the set of all the unique line configurations. Some representative line configurations are illustrated in Fig.~\ref{fig:figone}.


The coarse-grained Hamiltonian of a dislocation line in the absence of external stress is
\begin{equation}\label{eq:hamiltonian}
\mathcal{H} = f_k \sum_{i=1}^{n} |u_{i+1}-u_{i}| = 2 r f_k,
\end{equation}
where $f_k(T)$ is half the free energy of formation of a kink pair \footnote{{Left and right kinks on a $\frac{1}{2}\left\langle 111 \right\rangle$ bcc screw dislocation have different formation energies.\cite{bulatov1997kink, ventelon2009atomistic} Using the average formation energy is not an approximation because kinks are only formed in pairs, provided periodic boundary conditions apply.}} and $r$ is the number of kink pairs on the line. In this representation a kink is described as a quasiparticle with a distinct free energy of formation $f_k(T)$, with the dependence on temperature originating from the coarse-grained atomic degrees of freedom \cite{swinburne2018unsupervised}.
This model is closely related to the discrete Gaussian model \cite{chui1976phase}.

\textit{Equilibrium properties.---}
The atomic lattice acts as a heat bath with temperature $T$. In the absence of applied stress, biasing the motion of the dislocation line, or in other words in the limit $\sigma = 0$, the microstates containing identical numbers of kinks are degenerate. The system equilibrates, with thermal fluctuations giving rise to a balanced rate of creation and annihilation of paired left and right kinks. Placing first $r$ left kinks on $n$ available sites, and subsequently $r$ right kinks on $n-r$ remaining sites, we find the canonical partition function
\begin{equation}\label{eq:partitioneq}
Z  = \sum_{r=0}^{\lfloor n/2 \rfloor} 
            \binom{n}{r} \binom{n-r}{r} 
            z^r
    = \hgeom{\frac{1}{2}-\frac{n}{2},-\frac{n}{2}; 1; 4z},
\end{equation}
where $z = \exp(-2\beta f_k)$, $\beta = (\kB T)^{-1}$ is the reciprocal temperature, $\kB$ is the Boltzmann constant, and $\lfloor x \rfloor$ is the floor function. The above summation result is exact, and is expressed in terms of a hypergeometric function $\hgeom{a,b;c;x}$ \cite{beukers2007gauss}. A similar partition function was previously investigated in the limiting case where left and right kinks were assumed to be indistinguishable \cite{swinburne2018kink}.

The exact solution expressed as a hypergeometric function is not conducive to further analysis. Hence in what follows we restrict the discussion to the kink-dominated mobility regime, where $z \ll 1$. In this limit, pertinent to virtually all the conditions encountered in experiment \cite{seeger2004}, the partition function is given by the expression
\begin{equation}\label{eq:apppartitioneq}
Z \approx I_0(2n\sqrt{z}) 
\quad(z\ll1)
\end{equation}
where $I_0(x)$ is the modified Bessel function of the first kind of order zero. We refer to Appendix~\ref{sec:appendixderivation} for the detailed derivation.

Since the system studied here has a finite size $n$, the entropy $S = \partial_T (\kB T \ln Z)$ is not necessarily extrinsic. This is particularly evident in the low temperature (LT) limit, where
\begin{equation}
S_\mathrm{LT} \overset{n \gg 1}{\approx} n^2 z \bigl[
        \kB  + 
         2 f_k(T)/T - 2 f_k'(T)\bigr]
         + \mathcal{O}(z^2).
\end{equation}
In the high temperature (HT) limit, entropy has the form
\begin{equation}
S_\mathrm{HT} \overset{n \gg 1}{\approx} 2n\sqrt{z} \bigl[
            \kB + f_k(T)/T - f_k'(T) \bigr]
            + \mathcal{O}(z).
\end{equation}
As a function of system size and temperature, entropy changes from being intrinsic $(S_\mathrm{LT} \propto n^2 z)$ at low temperature to being extrinsic $(S_\mathrm{HT} \propto n \sqrt{z})$ at sufficiently high temperature. Furthermore, the argument of the exponential function, dominating the variation of entropy as a function of temperature, effectively halves in the high $T$ limit, from $2 f_k$ (where $S\sim z=\exp (-2\beta f_k)$) at low temperature to $f_k$ (where $S\sim \sqrt{z}=\exp(-\beta f_k)$) at high temperature, in agreement with Refs.~\cite{swinburne2018kink,hirth1982theory, po2016phenomenological}.

In what follows, we define the critical temperature $T^*$ as a temperature of a transition between the intrinsic and extrinsic regimes for a given dislocation segment size $n$. To illustrate the nature of this transition, consider for example the mean number of kink pairs $\left\langle r\right\rangle$ on the dislocation line
\begin{equation}
\left\langle r\right\rangle 
    =  -\beta^{-1}\frac{\partial}{\partial f_k} \ln Z 
    = 2n\sqrt{z}\, \frac{I_1(2n\sqrt{z})} {I_0(2n\sqrt{z})},
\end{equation}
where $I_1(x)$ is the modified Bessel function of the first kind of order one. Similarly to entropy, the expression for $\ln \left\langle r\right\rangle$ exhibits a change of slope as a function of  reciprocal temperature.  $\beta ^*= (k_B T^*)^{-1}$ corresponds to the point where the change of slope is maximum:
\begin{equation}\label{eq:root}
0 = \left.\frac{\partial^3}{\partial \beta^3} 
    \ln \left\langle r\right\rangle\right|_{\beta = \beta^*}.
\end{equation}
The value of $T^*$ defined by this equation depends both on the length of the dislocation segment $n$ and the free energy of formation of a kink pair $2f_k$. The resulting expression for the root of equation \eqref{eq:root} does not admit an analytical solution as it depends on the free energy function $f_k(T)$. Approximating the free energy of formation of a kink pair by a linear function of temperature $f_k(T) \approx a_0 + a_1 T$, which is motivated by atomistic free energy simulations in tungsten \cite{swinburne2018unsupervised}, we find that the critical temperature satisfies the following implicit equation
\begin{equation}\label{eq:ntransition}
 n = \xi\, \exp\!\left(\frac{f_k(T^*)}{\kB T^*}\right),
\end{equation}
where $\xi = 0.95483$ is a numerical constant.

\begin{figure}[t]
\includegraphics[width=.48\columnwidth]{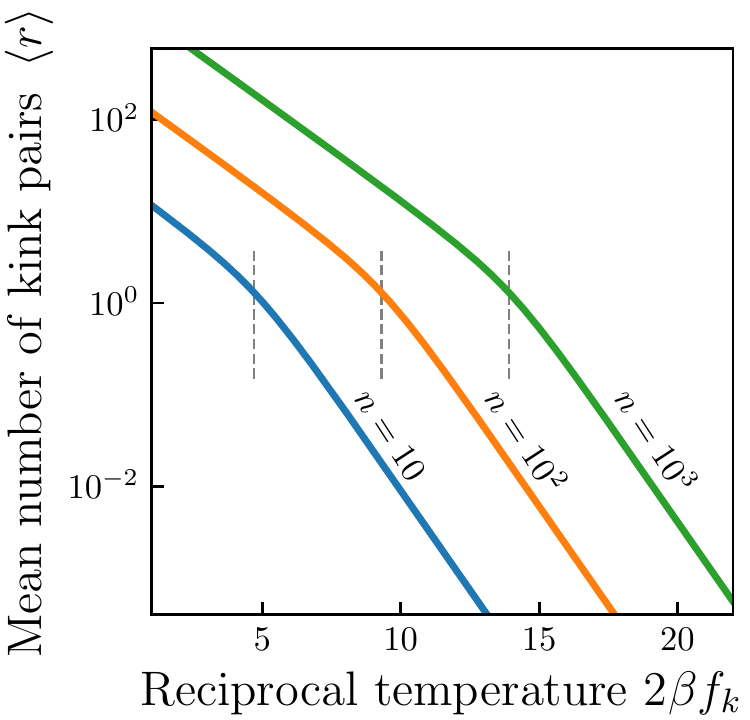}
\hspace{.01\columnwidth}
\includegraphics[width=.48\columnwidth]{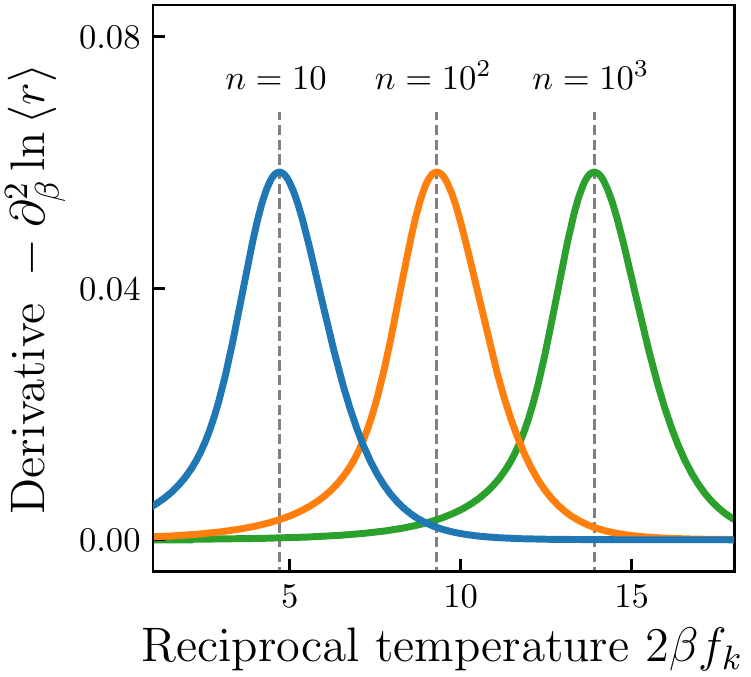}
\caption{Left: The average expected number of kink pairs $\left\langle r \right\rangle$ computed as a function of the scaled reciprocal temperature for dislocation segment lengths $n=10$, $100$, and $1000$. Right: the peak of the second-order derivative $-\partial^2_\beta \ln \left\langle r \right\rangle$ marks the critical temperature (dashed lines), see Eq.~\eqref{eq:ntransition}.}
\label{fig:figtwo} 
\end{figure}

The Arrhenius plot in Fig.~\ref{fig:figtwo} shows the dislocation segment size scaling of the transition. We note that this critical transition is not a conventional thermodynamic transition, as it occurs in a finite size system. A related \textit{apparent roughening} transition is also found in the sine-Gordon model \cite{ares2003apparent}, which is a continuum equivalent of the discrete Gaussian model \cite{baskaran1984equivalence}.

\

\textit{Non-equilibrium steady state.---}
The introduction of bias in the form of applied shear stress drives the dislocation line out of equilibrium. The rate of nucleation and propagation of kinks in the direction favoured by the applied bias is higher, while the processes occurring against the bias are suppressed. The presence of bias eventually gives rise to the steady-state drift of the dislocation line. With no loss of generality we restrict the following discussion to the case $\sigma >0$.

To evaluate the mean glide velocity, one needs to know the probability of occurrence of any microstate and its associated rate of escape into the connected microstates. We describe the kinetics of the line as a sequence of fundamental reactions, where each involves moving just one line segment at a time, either in the positive ($u_i \rightarrow u_i+1$) or negative ($u_i \rightarrow u_i-1$) sense with respect to the biasing direction. Each positive or negative reaction thereby changes the mean position of the line by $\pm h/n$. The reaction rates are distinguished by whether the reaction results in motion of a kink or creation of a kink pair:
\begin{enumerate}
    \item If the reaction $u_i \rightarrow u_i \pm 1$ increases the number of kinks on the line, a kink pair is created with rate $k_c^\pm$.
    \item If the reaction $u_i \rightarrow u_i \pm 1$ conserves or reduces the number of kinks on the line, a kink is moved with a rate $k_m^\pm$.
\end{enumerate}
For the sake of simplicity we assumed that the process of a kink pair annihilation is equivalent to a right kink moving to a site containing a left kink, or the other way around.

We begin by expressing the non-equilibrium dynamics of the system in terms of a Markov chain \cite{Markov}, as this will allow to systematically introduce the approximations required to arrive at an analytical expression for drift velocity. Let the state vector $(\vec{\pi}_m)_i$ contain the relative number of visits to microstate $C_i$ after $m$ reactions. The normalization $\sum_{i} (\vec{\pi}_m)_i = 1$ applies because no sinks or sources are present.
Depending on how the kinks are distributed in $C_i$, a microstate may reach a connected microstate $C_j$ through one of the fundamental reactions with rate $k_{ij}$. Defining $k_i = \sum_j k_{ij}$ as the total escape rate out of $C_i$, the transition matrix is obtained as $(\mathrm{P})_{ij} = k_{ij}/k_j$. The state vector is then iterated according to:
\begin{equation}
\vec{\pi}_{m+1} = \vec{\pi}_m \cdot \mathrm{P}
\end{equation}
The state vector converges independently of the initial value $\vec{\pi}_0$ to the same steady state $\lim_{m\rightarrow\infty} \vec{\pi}_m = \vec{\pi}$. Expectation values $\left\langle x \right\rangle$ in the steady state are obtained by the mean outcome $x_i$ weighted by the average time $\pi_i k_i^{-1}$ spent in a given microstate :
\begin{equation}\label{eq:expectationvalue}
\left\langle x \right\rangle 
        = \sum_i \frac{x_i \pi_i k_i^{-1}}{\sum_j \pi_j k_j^{-1}} = \sum_i p_i x_i,
\end{equation}
where $p_i$ is the probability for the system to be in microstate $C_i$.

The outcome for velocity $v_i$ of a microstate $C_i$ is obtained as the net balance between escape rates associated with reactions moving the line in the positive or negative sense with respect to the direction of applied stress
\begin{equation}\label{eq:velocityoutcome}
v_i = \frac{h}{n}\left(
    n_{i,c}^+ k_c^+ - n_{i,c}^- k_c^- + n_{i,m}^+ k_m^+ - n_{i,m}^- k_m^- \right),
\end{equation}
where $n_{i,\varepsilon}^\pm$ is the number of microstates connected to $C_i$ that can be reached by a fundamental reaction occurring at rate $k_{\varepsilon}^\pm$.

The transition matrix can only be built explicitly for small systems as the number of microstates scales exponentially with $n$. An exact solution for the drift velocity is consequently unfeasible. Instead we introduce a series of approximations that eventually lead to an analytical solution. 

First, we reduce the dimension of the state vector by partitioning microstates with similar probability of occurrence into non-overlapping groups $P_\alpha$, each containing a certain number of microstates $N_\alpha$. Next, we assume that the partitioning was chosen such that all the microstates $C_i \in P_\alpha$ in a group occur with approximately equal probability $p_i \approx N_\alpha^{-1}p_\alpha$, namely
\begin{equation}\label{eq:avgoutcome0}
\left\langle x \right\rangle = \sum_i p_i x_i \approx 
    \sum_{P_\alpha} N_\alpha^{-1} p_\alpha \sum_{i\in P_\alpha} x_i =
    \sum_{P_\alpha} p_\alpha \overline{x}_\alpha,
\end{equation}
where 
\begin{equation}\label{eq:avgoutcome}
\overline{x}_\alpha = N_\alpha^{-1} \sum_{i \in  P_\alpha} x_i
\end{equation}
is the group-averaged outcome. 

In practice, we choose to put together all the microstates with the same number of kink pairs $r$ into a group $P_r$. This is motivated by the fact that the rate of kink motion is much higher than the rate of kink creation, allowing the system to rapidly explore microstates with the same kink number between the reactions leading to the creation of kink pairs. The description could be improved by further grouping the states by area $A$, but this results in a severe combinatorial challenge \footnote{Determining the group size $N_{r,A}$ is equivalent to the lattice enumeration problem \cite{bona2015handbook} of finding the number of grand Motzkin paths from $(0,0)$ to $(n,0)$ weighted by up (or down) steps and area. We refer to some examples here \cite{owczarek2009exact, owczarek2010exact}. This is also required to determine the total number of reactions of each $(r,A)$ group into the connected groups.}.
%
In the limit where the bias is relatively weak, the occupation probability $p_r$ of a group $P_r$ will be very close to the equilibrium occupation probability $p^{\rm eq}_r$.
Equating the identity $Z=\sum_{r=0}^{\infty}p^{\rm eq}_r$ with the expansion 
$Z=\sum_{r=0}^{\infty} {n^{2r}}z^r/{(r!)^2}$ of Eq.~\eqref{eq:apppartitioneq}, we find
\begin{equation}\label{eq:expectationequilibrium}
\left\langle x \right\rangle 
    \approx \left\langle x \right\rangle_\mathrm{eq} 
    = \frac{1}{I_0(2n\sqrt{\tilde{z}})}
        \sum_{r = 0}^{\infty} \frac{n^{2r}}{(r!)^2} \tilde{z}^r\, \overline{x}_r  ,
\end{equation}
with $z$ replaced by the normalized average rate of creation of kink pairs $\tilde{z} = (k_c^+ + k_c^-)/(2 k_0)$ to approximately account for the biased creation of kink pairs in the preferred direction, with $k_0 = \left. k_c^\pm \right|_{\sigma,T=0}$. Note that the linear-response formalism is recovered by choosing $\tilde{z} = z$. In practice we find that the average rate gives an improved agreement at high stresses, see Appendix for a comparison. 

Following Eq.~\eqref{eq:avgoutcome}, we proceed to determine the group-averaged line velocity $\overline{v}_r$. To achieve this, we need to evaluate the total number of configurationally permitted escape reactions from a set of microstates in $P_r$. The resulting combinatorial problem can be solved exactly, provided that we restrict a site on the line to accommodate at most one kink. The exact solution is possible, it involves hypergeometric functions and is presented in the Appendix for the sake of completeness.  Here we continue with one final approximation. Bearing in mind that the approximations taken so far are valid in the kink-dominated regime $z \ll 1$, where the kink density is low \cite{seeger2004}, the number of microstates where kinks overlap is very small in comparison with the number of microstates where kinks do not overlap. Hence the number of microstates connected to a given group $P_r$ can be well approximated by considering every site to be free for kink nucleation $(n_{r,c}^\pm \approx n)$, and every kink to be free to move in either direction along the line $(n_{r,m}^\pm \approx 2r)$. Using this argument, we arrive at an expression for the group-average velocity in the form
\begin{equation}\label{eq:velocityoutcome2}
\overline{v}_r \approx \frac{h}{N_r} \left[
    n (k_c^+ - k_c^-) + 2 r (k_m^+ - k_m^-) 
    \right],
\end{equation}
where $N_{r} = n^{2r}/(r!)^2$ is the number of microstates containing $r$ kink pairs. Substituting \eqref{eq:velocityoutcome2} into \eqref{eq:expectationequilibrium} and expressing the sums as Bessel functions, we arrive at the central result of this manuscript, the approximate analytical expression for the glide velocity:
\begin{equation}\label{eq:glidevelocity}
 \left\langle v \right\rangle_\mathrm{eq} \approx 
    2 h \sqrt{\tilde{z}}(k_m^+ - k_m^-) 
    \frac{I_1[2n\sqrt{\tilde{z}}]}
    {I_0[2n\sqrt{\tilde{z}}]}.
\end{equation}
The term associated with the formation of kink pairs is not included as it is negligible in the kink-dominated regime $z \ll 1$. The assumptions leading to the derivation of the drift velocity are only strictly valid in the limit $z \ll 1$, pertinent to the overwhelming majority of experimental conditions \cite{seeger2004}. Leaving out the kink pair formation term has a simple but clear benefit of extending the single kink regime in the drift velocity without the onset of saturation effects at $z \sim 1$ and beyond
\begin{equation}
\lim_{\tilde{z} \rightarrow \infty}  \left\langle v \right\rangle_\mathrm{eq} = h k_0 \exp(-\beta f_k),
\end{equation}
which allows the velocity law to be manually extended at any point by interpolating it to other expressions describing dislocation mobility in the viscous drag or high-speed regime \cite{po2016phenomenological}. In the Appendix~\ref{sec:appendixexact} we present an expression for glide velocity derived using the exact combinatorics of non-overlapping kinks, and demonstrate that the approximate expression \eqref{eq:glidevelocity} is entirely consistent with the exact approach in the kink-dominated mobility regime.

\textit{Comparison to kMC.---}
We implemented a rejection-free kMC algorithm \cite{voter2007introduction} to propagate the line and extract the mean glide velocity. While kMC does not provide closed form analytical expressions for the expectation values, it serves as a reference for the approximate analytical solution $\eqref{eq:glidevelocity}$. We used the following model rates for the fundamental reactions:
\begin{equation}
\begin{aligned}
k_c^\pm &= k_0 \exp\left[-2\beta f_k(\pm\sigma, T)\right] \\
k_a^\pm &= k_0 \\
k_m^\pm &= k_0 \pm b \sigma/(2\gamma),
\end{aligned}
\end{equation}
and an expression for the free energy of a kink from Ref. \cite{swinburne2018kink}
\begin{equation}\label{eq:freeenergy}
f_k(\sigma, T) = U_k \left(
    1 - \frac{T}{T_\mathrm{ath}} - \frac{\sigma/\sigma_\mathrm{P}}{1-T/T_\mathrm{ath}}\right).
\end{equation}
A $\frac{1}{2}\left\langle 111 \right\rangle$ screw dislocation in iron is described by the following set of atomistically obtained parameters $b = \SI{2.47}{\Ang}$, $U_k = \SI{0.33}{\eV}$ \cite{gordon2011screw}, $T_\mathrm{ath} = \SI{700}{\kelvin}$ \cite{gilbert2013free}, $\sigma_\mathrm{P} = \SI{900}{\MPa}$ \cite{proville2012quantum}, and kink dissipation $\gamma = \SI{1.83}{m_u \ps^{-1}}$ \cite{swinburne2013theory}. Rates $k_m^\pm$ are derived by equating the mean velocity of a kink under shear stress \cite{swinburne2013theory} $\left\langle v \right\rangle = b h\sigma/\gamma$ with the net stochastic drift $\left\langle v \right\rangle = h (k_m^+ - k_m^-)$, where $k_m^\pm \sim k_0 (1 \pm \alpha\sigma)$, and subsequently solving for $\alpha$. Estimating $k_0$ from the Debye frequency $\sim \SI{10}{\ps^{-1}}$ gives $k_m^\pm = k_0 (1 \pm \SI{0.004}{\MPa^{-1}}\sigma)$. We assume that the energy barrier to kink pair formation is equal to the free energy of formation, which is consistent with the minimum energy pathways for kink formation obtained from atomistic simulations \cite{gordon2011screw, stukowski2015thermally, marinica2013interatomic, swinburne2018unsupervised}.

\begin{figure}[t]
\includegraphics[width=\columnwidth]{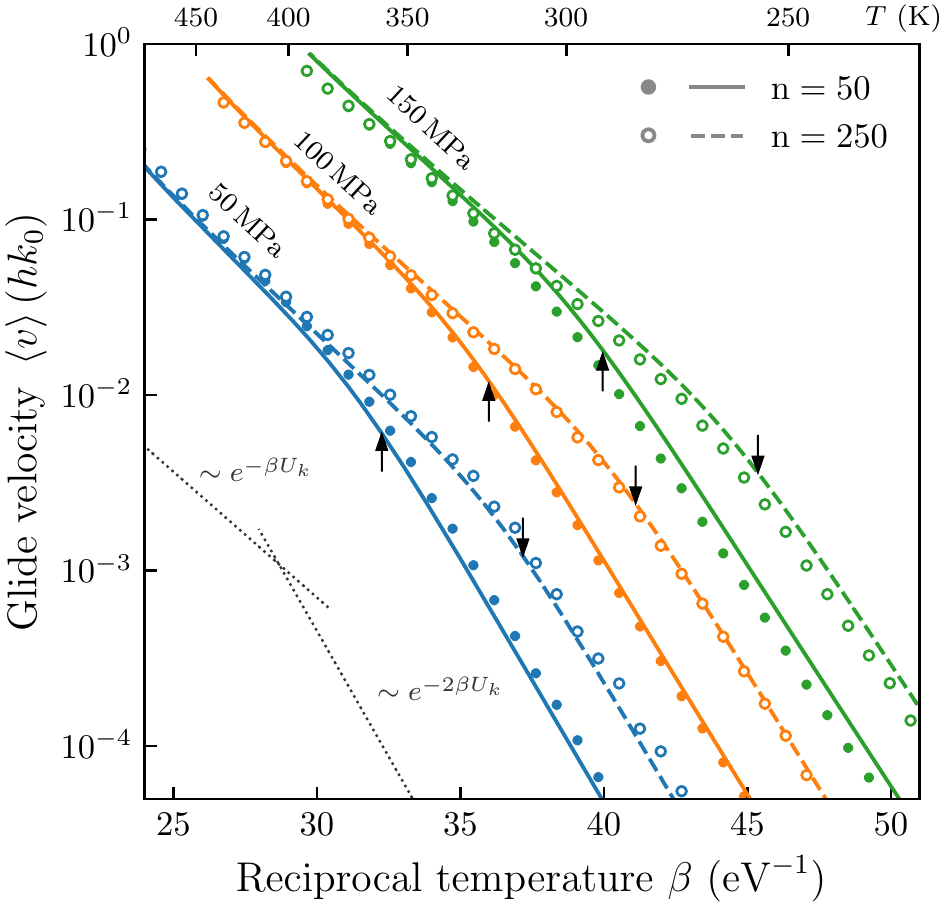}
\caption{Glide velocity of a segment of a $\frac{1}{2}\left\langle 111 \right\rangle$ screw dislocation in bcc iron with length $L=nb$ under shear stress. Analytical velocities (lines) from Eq.~\eqref{eq:glidevelocity} are consistent with velocities from kMC simulations (dots). Arrows mark the  predicted values of the critical temperature at which the activation energy changes from a kink pair $2f_k$ to a single kink $f_k$. Dotted lines show Arrhenius rates for single kink ($U_k$) and kink pair ($2U_k$) activation energies to guide the eye.}
\label{fig:figthree} 
\end{figure}

We ran kMC simulations for two dislocation segment lengths $n=50$ and $250$ for a range of stresses and temperatures for which $f_k > 0$, see Fig.~\ref{fig:figthree}. Statistical uncertainties in the velocities obtained by kMC simulations are below \SI{3}{\percent}. The analytically computed velocities \eqref{eq:glidevelocity} are in quantitative agreement with velocities derived from kMC simulations. An estimate for the critical temperature for a gliding screw dislocation is obtained by analogy with the equilibrium case \eqref{eq:ntransition} using the condition $n = \xi/\sqrt{\tilde{z}^*}$. The largest discrepancy is found for $\sigma=\SI{150}{\MPa}$ at low temperature, where the analytical expression for velocity overestimates the velocity found in kMC simulations by about \SI{50}{\percent}, which is expected since the linear response approximation underlying the derivation assumes the low stress limit.

\begin{figure}[t]
\includegraphics[width=\columnwidth]{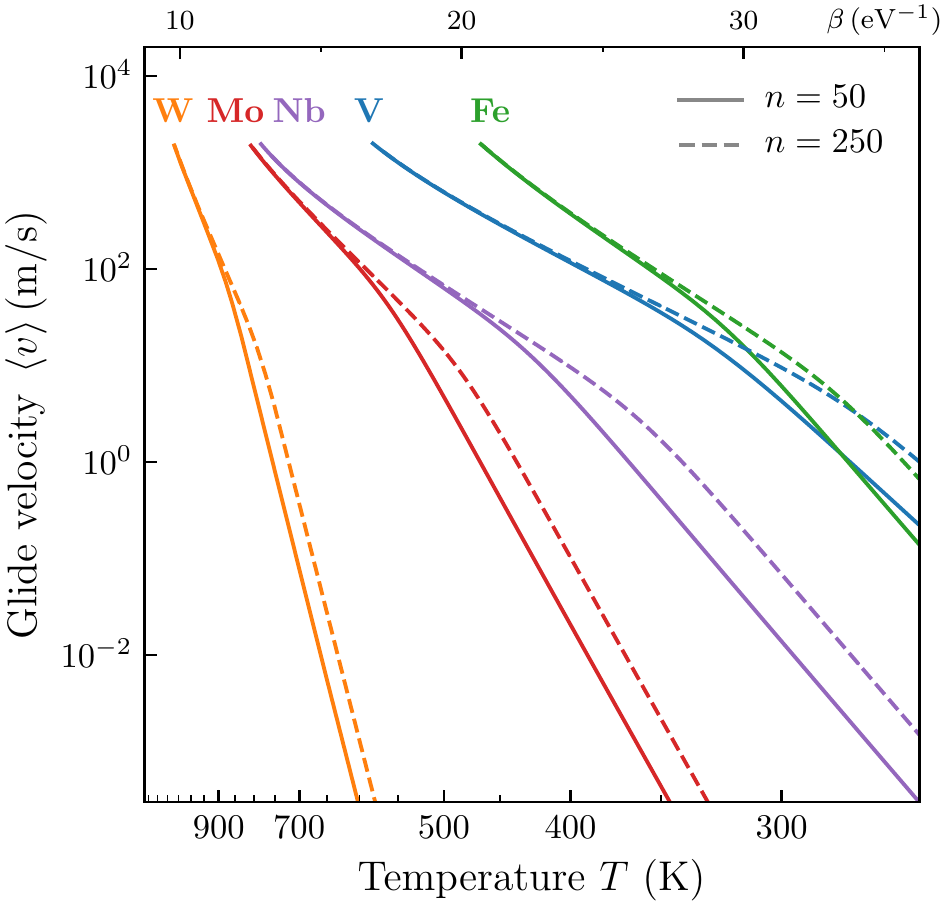}
\caption{Glide velocity of a $\frac{1}{2}\left\langle 111 \right\rangle$ screw dislocation segment with length $L=nb$ under a resolved shear stress of \SI{50}{\MPa} in a selection of bcc materials, predicted using Eq.~\eqref{eq:glidevelocity} with parameters from Tab.~\ref{tab:partable} obtained from experiments. Velocities larger than \SI{2e3}{m/s} are not plotted since the model is not suited to describe dislocation velocities approaching the speed of sound in the material.}
\label{fig:figfour} 
\end{figure}

Using the analytical expression \eqref{eq:glidevelocity}, it is possible to predict the velocity of a gliding screw dislocation for any bcc material as a function of temperature, stress, and length; provided a model for the free energy of formation of a kink is available. Parameters $U_k$ and $\sigma_\mathrm{P}$ are readily available from experimental studies \cite{seeger2004,giannattasio2007empirical}, while we infer $T_\mathrm{ath}$ and $k_m^\pm$ by scaling the parameters for iron, see Tab.~\ref{tab:partable}. The glide velocities of a screw dislocation segment under a resolved shear stress of $\SI{50}{\MPa}$ in bcc metals W, Mo, Nb, V, and Fe, are shown in Fig.~\ref{fig:figfour}, assuming attempt rate $k_0 = \SI{10}{\ps^{-1}}$ and kink height $h=\sqrt{2/3}a$.

\begin{table}[t]
\caption{Materials parameters for kink formation energy \eqref{eq:freeenergy}. Athermal temperatures and kink mobilities are estimated by scaling the value for iron by the ratio of melting temperatures ($T_\mathrm{ath} = \SI{700}{\K}\times T_\mathrm{melt}^\mathrm{mat}/T_\mathrm{melt}^\mathrm{Fe}$) and shear moduli ($\alpha = \SI{0.004}{\MPa^{-1}}\times \mu^\mathrm{mat}/\mu^\mathrm{Fe}$), respectively.
\label{tab:partable}}
\begin{ruledtabular}
  \begin{tabular}{ccccc}
  	material & $U_k$ (\si{\eV}) & $\sigma_\mathrm{P}$ (\si{\MPa}) & $T_\mathrm{ath}$ (\si{\K}) & $\alpha$ (\si{\MPa^{-1}}) \\ \hline
	V     &	0.27 \cite{giannattasio2007empirical}  &   350 \cite{suzuki1999plastic}%
	            &  840 & 0.003 \\
	W     &	1.05 \cite{giannattasio2007empirical}  &   960 \cite{kamimura2013experimental}%
	            & 1400 & 0.009 \\
	Fe    &	0.33 \cite{giannattasio2007empirical}  &   390 \cite{kamimura2013experimental}%
	            &  700 & 0.004 \\
	Mo    &	0.49 \cite{giannattasio2007empirical}  &   730 \cite{kamimura2013experimental}%
	            & 1100 & 0.007 \\
	Nb    &	0.34 \cite{seeger2006slip}             &   415 \cite{kamimura2013experimental}%
	            & 1100 & 0.002 \\
  \end{tabular}	
\end{ruledtabular}
\end{table}

\textit{Concluding remarks.---} 
We derived an analytical expression for the thermally activated glide velocity of a screw dislocation segment of arbitrary length in a bcc metal, driven by external stress. The model accounts for formation energies and rates of fundamental reactions involving kinks, which are accessible to atomistic simulations \cite{rodney2009stress, swinburne2013theory,  proville2013prediction, maresca2018screw}, and as such is generally transferable across bcc metals. We show that the activation energy for dislocation glide is halved at high temperature, as the system crosses a critical transition into the thermodynamic limit. The mobility law generalizes past formulations of kink-limited motion of screw dislocations. We note that dislocation mobility enters the viscous drag regime \cite{gilbert2011stress, po2016phenomenological} for $f_k \lesssim 0$, which lies outside the scope of this work.

Our model shows that the activation energy for plastic deformation can indeed be lowered to $f_k$, however, it also suggests that the activation energy should be closer to $2f_k$ in irradiated microstructures, as the mean dislocation length is reduced by radiation defects acting as dislocation pinning points \cite{swinburne2018kink}. The dynamics of screw dislocations terminating at features often found in realistic microstructure, such as junctions, grain boundaries, or surfaces, is a largely unexplored area. We hence emphasize the need for atomistic studies of kinetics of such systems in order to fully classify the rate-limiting mechanisms contributing to plastic deformation of bcc metals.

\

\section*{Acknowledgments}

\begin{small}
\noindent
This work has been carried out within the framework of the EUROfusion Consortium and has received funding from the Euratom research and training programme 2014-2018 and 2019-2020 under Grant Agreements No. 633053 and No. 755039. Also, it has been partially funded by the RCUK Energy Programme (Grant No. EP/T012250/1). The views and opinions expressed herein do not necessarily reflect those of the European Commission. Work at Los Alamos National Laboratory was supported  by the Laboratory Directed Research and Development program under project 20190034ER. Los Alamos National Laboratory is operated by Triad National Security LLC, for the National Nuclear Security administration of the  U.S. DOE under Contract No. 89233218CNA0000001.

\end{small}

\appendix

\section{Derivation of the approximate partition function}\label{sec:appendixderivation}

The canonical partition function \eqref{eq:partitioneq} is transformed to a double summation in anticipation of the following manipulations:
\begin{equation}
Z = \sum _{r=0}^{\lfloor n/2 \rfloor} 
  \sum _{s=0}^{2r} \binom{n}{2r} \binom{2r}{s} \delta_{rs}z^r.
\end{equation}
Using the integral representation for the Kronecker delta symbol 
\begin{equation}
\delta_{rs} = \frac{1}{2\pi} \int \limits _0 ^{2\pi} \dint{\xi}\, e^{i(r-s)\xi}
\end{equation}
and the expression
\begin{equation}
\sum_{i=0}^{\lfloor n/2 \rfloor}  \binom{n}{2i} x^{2i}  
		= \frac{1}{2}(1 + x)^n + \frac{1}{2}(1 - x)^n,
\end{equation}
 we write
\begin{align}
Z
  &= \frac{1}{2} \int \limits_0 ^{2\pi} \dint{\xi}
  \sum _{r=0}^{\lfloor n/2 \rfloor} 
    \binom{n}{2r} 
    z^r e^{ir\xi}
  \sum _{s=0}^{2r}
    \binom{2r}{s} 
    e^{-is\xi}
\nonumber\\
  &=
  \frac{1}{2} \int \limits_0 ^{2\pi} \dint{\xi}
  \sum _{r=0}^{\lfloor n/2 \rfloor} 
        \binom{n}{2r} z^r e^{ir\xi}
  \left(1+e^{-i\xi}\right)^{2r}
  \nonumber\\
  &=
  \frac{1}{2} \int \limits_0 ^{2\pi} \dint{\xi}
  \sum _{r=0}^{\lfloor n/2 \rfloor} 
  \binom{n}{2r} 
  \left[2\sqrt{z}\cos\left(\xi/2\right)\right]^{2r}
  \nonumber\\
  &=
  \frac{1}{2} \int \limits_0 ^{2\pi} \dint{\xi}
  \left[ 1+2\sqrt{z}\cos\xi \right]^{ n} \label{eq:2f1int}\\
  &=
  \frac{1}{2} \int \limits_0 ^{2\pi} \dint{\xi}
  \exp\left[ n
  \ln\left(1+2\sqrt{z}\cos\xi\right)
  \right].
\end{align}
Since in the kink-dominated regime we have $\sqrt{z}\ll 1$, the logarithmic term in the above equation can be simplified using a Taylor expansion and neglecting the terms quadratic in $\sqrt{z}$, leading to the expression
\begin{equation}
Z \approx I_0 (2 n \sqrt{z}),
\end{equation}
where $I_0(x)$ is the modified Bessel function of the first kind of order zero. We note that the exact solution in terms of the hypergeometric function is identified as the integral \eqref{eq:2f1int}.

\section{Statistics of the exact partition function and comparison to kMC}\label{sec:appendixexact}

In the manuscript we defined the critical value $T^*$ as the point where the change of slope of $\ln \left\langle r\right\rangle$ is maximal under the approximation that the free energy of formation of a kink pair is given by a linear function of temperature. The same procedure is repeated for the exact partition function \eqref{eq:partitioneq} with the mean number of kink pairs
\begin{equation}
\begin{aligned}
\left\langle r\right\rangle &= 2 n (n-1)\, \frac{
        \hgeom{\frac{3}{2}-\frac{n}{2}, 1-\frac{n}{2}; 2; 4z}
    }{
        \hgeom{\frac{1}{2}-\frac{n}{2},-\frac{n}{2};1;4z}
    }.
\end{aligned}
\end{equation}
This time the stationary point depends on system size $n$, therefore we investigate the equation numerically on the interval $n \in [10^1,10^5]$, leading to the following implicit equation:
\begin{equation}\label{eq:ntransition_app}
 n = \xi(n)\, \exp\!\left(\frac{f_k(T^*)}{\kB T^*}\right),
\end{equation}
where $\xi(n)$ in the limit $n\gg1$ is well approximated by the series
\begin{equation}
\xi(n) = 0.9548 + \frac{1.2938}{n} - \frac{7.1232}{n^2}+\dots 
\end{equation}
It is remarkable that for large systems, where $n\rightarrow \infty$, the scaling function $\xi(n)$ becomes identical to the numerical constant $\xi = 0.9548$ determined for the approximate partition function. In the exact partition function we permit up to one kink per site, and hence do not include microstates with overlapping kinks. As the system size becomes large, for a given number of kink pairs $r$, the number of microstates with overlapping kinks becomes negligible compared to the number of microstates with non-overlapping kinks, which reaffirms that the exact and approximate partition functions are equivalent at low kink densities.

The drift velocity for the system described by the exact partition function is determined in the linear response approximation
\begin{equation}\label{eq:expectationequilibrium_app}
 \left\langle v\right\rangle_\mathrm{eq} 
    = Z^{-1} \sum_{r = 0}^{\lfloor n/2 \rfloor}  \binom{n}{r} \binom{n-r}{r} \tilde{z}^r \overline{v}_r,
\end{equation}
where $Z= \hgeom{\tfrac{1}{2}-\tfrac{n}{2},-\tfrac{n}{2}; 1; 4\tilde{z}}$ is the exact partition function. As before, we put together all the microstates with the same number of kink pairs into a group $P_r$. 

It remains to determine the group-averaged line velocity $\overline{v}_r$ \eqref{eq:avgoutcome}, which requires knowledge of the total number of microstates connected to each group $n_{r,\varepsilon}^\pm = \sum_{i\in P_r} n_{i,\varepsilon}^\pm$ for each fundamental reaction or process. This combinatorial problem is exactly solvable for non-overlapping kinks. We begin by considering the process of kink pair annihilation, here explicitly included for completeness with a rate consistent with kink motion $k_a^\pm = k_m^\pm$. Any microstate $C_i$ with non-overlapping kinks can equivalently be described by a sequence of horizontal steps \textsc{h}, left kink steps \textsc{l}, and right kink steps \textsc{r}, with the number of \textsc{l} and \textsc{r} steps each being equal to $r$. A process of kink pair annihilation advancing the line towards the biasing direction is described by a subsequence (\textsc{l},\textsc{r}) in position $(i, i+1)$ modulo $n$ turning into the subsequence (\textsc{h},\textsc{h}), with the rest of the line remaining unchanged. There are $n$ ways to choose $i$, and of the remaining $(n-2)$ sites we need to choose $(r-1)$ steps of \textsc{l} and \textsc{r} each, giving the total number of annihilation processes in $P_r$ as
\begin{equation}
n_{r,a}^\pm = \sum_{i\in P_r} n_{i,a}^\pm = n \binom{n-2}{r-1} \binom{n-r-1}{r-1},
\end{equation}
where we used $n_{i,\varepsilon}^+ = n_{i,\varepsilon}^-$ due to symmetry. Next, we recognize that the total number of pair annihilation processes in group $P_r$ corresponds precisely to the total number of pair creation processes from $P_{r-1}$, leading to $n_{r,a}^\pm = n_{r-1,c}^\mp$. We use a similar argument to determine the number of processes of kink motion advancing the line towards the biasing direction, with a process described by either subsequence (\textsc{r}, \textsc{h}) turning into (\textsc{h}, \textsc{r}), or (\textsc{h}, \textsc{l}) turning into (\textsc{l}, \textsc{h}). The resulting group-averaged velocity equals 
\begin{equation}\label{eq:vpartition}
\begin{aligned}
\overline{v}_r 
= \frac{h}{N_r} \bigg[
      & \binom{n-2}{r}    \binom{n-r-2}{r}   (k_c^+ - k_c^-) \\
    + & 2\binom{n-2}{r-1} \binom{n-r-1}{r}   (k_m^+ - k_m^-) \\
    + & \binom{n-2}{r-1}  \binom{n-r-1}{r-1} (k_a^+ - k_a^-) 
    \bigg],
\end{aligned}
\end{equation}
where $N_{r} = \binom{n}{r}\binom{n-r}{r}$ is the number of microstates containing $r$ kink pairs. Substituting \eqref{eq:vpartition} into \eqref{eq:expectationequilibrium_app}, we arrive at the analytical expression for the glide velocity:
\begin{equation}\label{eq:driftexact}
\begin{aligned}
 \left\langle v \right\rangle & {}_\mathrm{eq} =
    \hgeom{\frac{1}{2}-\frac{n}{2},-\frac{n}{2}; 1; 4\tilde{z}}^{-1} \\
    \times h \bigg[
       & (k_c^+ - k_c^-) \,
        \hgeom{\tfrac{3}{2}-\tfrac{n}{2},1-\tfrac{n}{2}; 1; 4 \tilde{z}} \\
    +  \tilde{z} &(k_a^+ - k_a^-)\,
        \hgeom{\tfrac{3}{2}-\tfrac{n}{2},1-\tfrac{n}{2}; 1; 4 \tilde{z}} \\
    +
        2\tilde{z} &(n-2)(k_m^+ - k_m^-)\, \hgeom{\tfrac{3}{2}-\tfrac{n}{2},2-\tfrac{n}{2}; 2; 4 \tilde{z}}
     \bigg].
\end{aligned}
\end{equation}
The above expression is \textit{exact} in the sense that the combinatorics of non-overlapping kinks is explicitly accounted for.

\begin{figure}[t]
\includegraphics[width=\columnwidth]{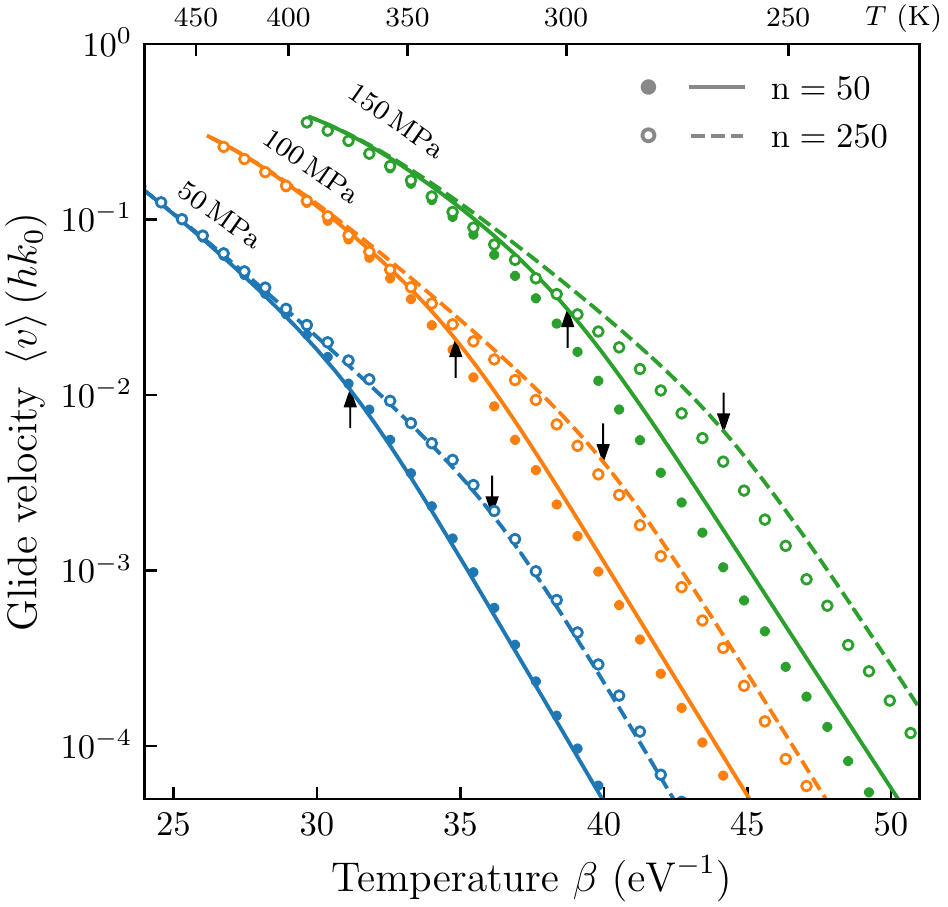}
\caption{Glide velocity of a segment of a $\frac{1}{2}\left\langle 111 \right\rangle$ screw dislocation in bcc iron with length $L=nb$ under shear stress. Analytical velocities (lines) following from the exact partition function \eqref{eq:driftexact} are consistent with velocities from kMC simulations with non-overlapping kinks (dots). See caption of Fig.~\eqref{fig:figthree} for more information.}
\label{fig:figthree_v2} 
\end{figure}

The derivation of the drift velocity \eqref{eq:driftexact} relies on the linear response approximation and the assumption that all microstates with the same number of kink pairs appear with equal probability. In contrast to the solution presented in the manuscript which is valid for $z \ll 1$, here we did not assume a low kink density with the caveat that kinks were not permitted to overlap. Consequently, at high temperature the line saturates with kinks, leading to an eventual suppression of kink creation and motion processes. In Fig.~\ref{fig:figthree_v2} we compare the analytical expression for drift velocity \eqref{eq:driftexact} to numerical simulation. Note that the kinetic Monte Carlo simulation here is restricted to only permit one kink per site, in accordance with the combinatorics of non-overlapping kinks. The onset of a saturation in velocity is apparent at high temperature, while at low temperature, where $z \ll 1$, the velocities are identical to the approximate solution \eqref{eq:glidevelocity}.

\bibliography{main.bbl}

\begin{thebibliography}{48}%
\makeatletter
\providecommand \@ifxundefined [1]{%
 \@ifx{#1\undefined}
}%
\providecommand \@ifnum [1]{%
 \ifnum #1\expandafter \@firstoftwo
 \else \expandafter \@secondoftwo
 \fi
}%
\providecommand \@ifx [1]{%
 \ifx #1\expandafter \@firstoftwo
 \else \expandafter \@secondoftwo
 \fi
}%
\providecommand \natexlab [1]{#1}%
\providecommand \enquote  [1]{``#1''}%
\providecommand \bibnamefont  [1]{#1}%
\providecommand \bibfnamefont [1]{#1}%
\providecommand \citenamefont [1]{#1}%
\providecommand \href@noop [0]{\@secondoftwo}%
\providecommand \href [0]{\begingroup \@sanitize@url \@href}%
\providecommand \@href[1]{\@@startlink{#1}\@@href}%
\providecommand \@@href[1]{\endgroup#1\@@endlink}%
\providecommand \@sanitize@url [0]{\catcode `\\12\catcode `\$12\catcode
  `\&12\catcode `\#12\catcode `\^12\catcode `\_12\catcode `\%12\relax}%
\providecommand \@@startlink[1]{}%
\providecommand \@@endlink[0]{}%
\providecommand \url  [0]{\begingroup\@sanitize@url \@url }%
\providecommand \@url [1]{\endgroup\@href {#1}{\urlprefix }}%
\providecommand \urlprefix  [0]{URL }%
\providecommand \Eprint [0]{\href }%
\providecommand \doibase [0]{https://doi.org/}%
\providecommand \selectlanguage [0]{\@gobble}%
\providecommand \bibinfo  [0]{\@secondoftwo}%
\providecommand \bibfield  [0]{\@secondoftwo}%
\providecommand \translation [1]{[#1]}%
\providecommand \BibitemOpen [0]{}%
\providecommand \bibitemStop [0]{}%
\providecommand \bibitemNoStop [0]{.\EOS\space}%
\providecommand \EOS [0]{\spacefactor3000\relax}%
\providecommand \BibitemShut  [1]{\csname bibitem#1\endcsname}%
\let\auto@bib@innerbib\@empty
\bibitem [{\citenamefont {Tanaka}\ \emph {et~al.}(2008)\citenamefont {Tanaka},
  \citenamefont {Tarleton},\ and\ \citenamefont {Roberts}}]{tanaka2008brittle}%
  \BibitemOpen
  \bibfield  {author} {\bibinfo {author} {\bibfnamefont {M.}~\bibnamefont
  {Tanaka}}, \bibinfo {author} {\bibfnamefont {E.}~\bibnamefont {Tarleton}},\
  and\ \bibinfo {author} {\bibfnamefont {S.~G.}\ \bibnamefont {Roberts}},\
  }\bibfield  {title} {\bibinfo {title} {The brittle--ductile transition in
  single-crystal iron},\ }\href@noop {} {\bibfield  {journal} {\bibinfo
  {journal} {Acta Materialia}\ }\textbf {\bibinfo {volume} {56}},\ \bibinfo
  {pages} {5123} (\bibinfo {year} {2008})}\BibitemShut {NoStop}%
\bibitem [{\citenamefont {Zinkle}(2012)}]{zinkle2012nuclear}%
  \BibitemOpen
  \bibfield  {author} {\bibinfo {author} {\bibfnamefont {S.~J.}\ \bibnamefont
  {Zinkle}},\ }\bibfield  {title} {\bibinfo {title} {Radiation-induced effects
  on microstructure},\ }in\ \href@noop {} {\emph {\bibinfo {booktitle}
  {Comprehensive Nuclear Materials}}},\ Vol.~\bibinfo {volume} {1},\ \bibinfo
  {editor} {edited by\ \bibinfo {editor} {\bibfnamefont {R.~J.~M.}\
  \bibnamefont {Konings}}, \bibinfo {editor} {\bibfnamefont {T.~R.}\
  \bibnamefont {Allen}}, \bibinfo {editor} {\bibfnamefont {R.~E.}\ \bibnamefont
  {Stoller}},\ and\ \bibinfo {editor} {\bibfnamefont {S.}~\bibnamefont
  {Yamanaka}}}\ (\bibinfo  {publisher} {Elsevier},\ \bibinfo {address}
  {Amsterdam},\ \bibinfo {year} {2012})\ Chap.~\bibinfo {chapter} {3}, pp.\
  \bibinfo {pages} {65--98}\BibitemShut {NoStop}%
\bibitem [{\citenamefont {Federici}\ \emph {et~al.}(2017)\citenamefont
  {Federici}, \citenamefont {Biel}, \citenamefont {Gilbert}, \citenamefont
  {Kemp}, \citenamefont {Taylor},\ and\ \citenamefont
  {Wenninger}}]{federici2017european}%
  \BibitemOpen
  \bibfield  {author} {\bibinfo {author} {\bibfnamefont {G.}~\bibnamefont
  {Federici}}, \bibinfo {author} {\bibfnamefont {W.}~\bibnamefont {Biel}},
  \bibinfo {author} {\bibfnamefont {M.~R.}\ \bibnamefont {Gilbert}}, \bibinfo
  {author} {\bibfnamefont {R.}~\bibnamefont {Kemp}}, \bibinfo {author}
  {\bibfnamefont {N.}~\bibnamefont {Taylor}},\ and\ \bibinfo {author}
  {\bibfnamefont {R.}~\bibnamefont {Wenninger}},\ }\bibfield  {title} {\bibinfo
  {title} {European {DEMO} design strategy and consequences for materials},\
  }\href@noop {} {\bibfield  {journal} {\bibinfo  {journal} {Nuclear Fusion}\
  }\textbf {\bibinfo {volume} {57}},\ \bibinfo {pages} {092002} (\bibinfo
  {year} {2017})}\BibitemShut {NoStop}%
\bibitem [{\citenamefont {Pintsuk}\ \emph {et~al.}(2019)\citenamefont
  {Pintsuk}, \citenamefont {Diegele}, \citenamefont {Dudarev}, \citenamefont
  {Gorley}, \citenamefont {Henry}, \citenamefont {Reiser},\ and\ \citenamefont
  {Rieth}}]{pintsuk2019european}%
  \BibitemOpen
  \bibfield  {author} {\bibinfo {author} {\bibfnamefont {G.}~\bibnamefont
  {Pintsuk}}, \bibinfo {author} {\bibfnamefont {E.}~\bibnamefont {Diegele}},
  \bibinfo {author} {\bibfnamefont {S.~L.}\ \bibnamefont {Dudarev}}, \bibinfo
  {author} {\bibfnamefont {M.}~\bibnamefont {Gorley}}, \bibinfo {author}
  {\bibfnamefont {J.}~\bibnamefont {Henry}}, \bibinfo {author} {\bibfnamefont
  {J.}~\bibnamefont {Reiser}},\ and\ \bibinfo {author} {\bibfnamefont
  {M.}~\bibnamefont {Rieth}},\ }\bibfield  {title} {\bibinfo {title} {European
  materials development: {Results} and perspective},\ }\href@noop {} {\bibfield
   {journal} {\bibinfo  {journal} {Fusion Engineering and Design}\ }\textbf
  {\bibinfo {volume} {146}},\ \bibinfo {pages} {1300} (\bibinfo {year}
  {2019})}\BibitemShut {NoStop}%
\bibitem [{\citenamefont {Brunner}(2000)}]{brunner2000comparison}%
  \BibitemOpen
  \bibfield  {author} {\bibinfo {author} {\bibfnamefont {D.}~\bibnamefont
  {Brunner}},\ }\bibfield  {title} {\bibinfo {title} {Comparison of flow-stress
  measurements on high-purity tungsten single crystals with the kink-pair
  theory},\ }\href@noop {} {\bibfield  {journal} {\bibinfo  {journal}
  {Materials Transactions, JIM}\ }\textbf {\bibinfo {volume} {41}},\ \bibinfo
  {pages} {152} (\bibinfo {year} {2000})}\BibitemShut {NoStop}%
\bibitem [{\citenamefont {Peierls}(1940)}]{peierls1940size}%
  \BibitemOpen
  \bibfield  {author} {\bibinfo {author} {\bibfnamefont {R.}~\bibnamefont
  {Peierls}},\ }\bibfield  {title} {\bibinfo {title} {The size of a
  dislocation},\ }\href@noop {} {\bibfield  {journal} {\bibinfo  {journal}
  {Proceedings of the Physical Society}\ }\textbf {\bibinfo {volume} {52}},\
  \bibinfo {pages} {34} (\bibinfo {year} {1940})}\BibitemShut {NoStop}%
\bibitem [{\citenamefont {Nabarro}(1947)}]{nabarro1947dislocations}%
  \BibitemOpen
  \bibfield  {author} {\bibinfo {author} {\bibfnamefont {F.~R.~N.}\
  \bibnamefont {Nabarro}},\ }\bibfield  {title} {\bibinfo {title} {Dislocations
  in a simple cubic lattice},\ }\href@noop {} {\bibfield  {journal} {\bibinfo
  {journal} {Proceedings of the Physical Society}\ }\textbf {\bibinfo {volume}
  {59}},\ \bibinfo {pages} {256} (\bibinfo {year} {1947})}\BibitemShut
  {NoStop}%
\bibitem [{\citenamefont {Braun}\ and\ \citenamefont
  {Kivshar}(1998)}]{braun1998nonlinear}%
  \BibitemOpen
  \bibfield  {author} {\bibinfo {author} {\bibfnamefont {O.~M.}\ \bibnamefont
  {Braun}}\ and\ \bibinfo {author} {\bibfnamefont {Y.~S.}\ \bibnamefont
  {Kivshar}},\ }\bibfield  {title} {\bibinfo {title} {Nonlinear dynamics of the
  {Frenkel}--{Kontorova} model},\ }\href@noop {} {\bibfield  {journal}
  {\bibinfo  {journal} {Physics Reports}\ }\textbf {\bibinfo {volume} {306}},\
  \bibinfo {pages} {1} (\bibinfo {year} {1998})}\BibitemShut {NoStop}%
\bibitem [{\citenamefont {Fitzgerald}(2016)}]{fitzgerald2016kink}%
  \BibitemOpen
  \bibfield  {author} {\bibinfo {author} {\bibfnamefont {S.~P.}\ \bibnamefont
  {Fitzgerald}},\ }\bibfield  {title} {\bibinfo {title} {Kink pair production
  and dislocation motion},\ }\href@noop {} {\bibfield  {journal} {\bibinfo
  {journal} {Scientific Reports}\ }\textbf {\bibinfo {volume} {6}},\ \bibinfo
  {pages} {39708} (\bibinfo {year} {2016})}\BibitemShut {NoStop}%
\bibitem [{\citenamefont {Swinburne}\ \emph {et~al.}(2013)\citenamefont
  {Swinburne}, \citenamefont {Dudarev}, \citenamefont {Fitzgerald},
  \citenamefont {Gilbert},\ and\ \citenamefont {Sutton}}]{swinburne2013theory}%
  \BibitemOpen
  \bibfield  {author} {\bibinfo {author} {\bibfnamefont {T.~D.}\ \bibnamefont
  {Swinburne}}, \bibinfo {author} {\bibfnamefont {S.~L.}\ \bibnamefont
  {Dudarev}}, \bibinfo {author} {\bibfnamefont {S.~P.}\ \bibnamefont
  {Fitzgerald}}, \bibinfo {author} {\bibfnamefont {M.~R.}\ \bibnamefont
  {Gilbert}},\ and\ \bibinfo {author} {\bibfnamefont {A.~P.}\ \bibnamefont
  {Sutton}},\ }\bibfield  {title} {\bibinfo {title} {Theory and simulation of
  the diffusion of kinks on dislocations in bcc metals},\ }\href@noop {}
  {\bibfield  {journal} {\bibinfo  {journal} {Physical Review B}\ }\textbf
  {\bibinfo {volume} {87}},\ \bibinfo {pages} {064108} (\bibinfo {year}
  {2013})}\BibitemShut {NoStop}%
\bibitem [{\citenamefont {Giannattasio}\ \emph {et~al.}(2007)\citenamefont
  {Giannattasio}, \citenamefont {Tanaka}, \citenamefont {Joseph},\ and\
  \citenamefont {Roberts}}]{giannattasio2007empirical}%
  \BibitemOpen
  \bibfield  {author} {\bibinfo {author} {\bibfnamefont {A.}~\bibnamefont
  {Giannattasio}}, \bibinfo {author} {\bibfnamefont {M.}~\bibnamefont
  {Tanaka}}, \bibinfo {author} {\bibfnamefont {T.~D.}\ \bibnamefont {Joseph}},\
  and\ \bibinfo {author} {\bibfnamefont {S.~G.}\ \bibnamefont {Roberts}},\
  }\bibfield  {title} {\bibinfo {title} {An empirical correlation between
  temperature and activation energy for brittle-to-ductile transitions in
  single-phase materials},\ }\href@noop {} {\bibfield  {journal} {\bibinfo
  {journal} {Physica Scripta}\ }\textbf {\bibinfo {volume} {2007}},\ \bibinfo
  {pages} {87} (\bibinfo {year} {2007})}\BibitemShut {NoStop}%
\bibitem [{\citenamefont {Abernethy}\ \emph {et~al.}(2019)\citenamefont
  {Abernethy}, \citenamefont {Gibson}, \citenamefont {Giannattasio},
  \citenamefont {Murphy}, \citenamefont {Wouters}, \citenamefont {Bradnam},
  \citenamefont {Packer}, \citenamefont {Gilbert}, \citenamefont {Klimenkov},
  \citenamefont {Rieth}, \citenamefont {Schneider}, \citenamefont {Hardie},
  \citenamefont {Roberts},\ and\ \citenamefont
  {Armstrong}}]{abernethy2019effects}%
  \BibitemOpen
  \bibfield  {author} {\bibinfo {author} {\bibfnamefont {R.~G.}\ \bibnamefont
  {Abernethy}}, \bibinfo {author} {\bibfnamefont {J.~S. K.-L.}\ \bibnamefont
  {Gibson}}, \bibinfo {author} {\bibfnamefont {A.}~\bibnamefont
  {Giannattasio}}, \bibinfo {author} {\bibfnamefont {J.~D.}\ \bibnamefont
  {Murphy}}, \bibinfo {author} {\bibfnamefont {O.}~\bibnamefont {Wouters}},
  \bibinfo {author} {\bibfnamefont {S.}~\bibnamefont {Bradnam}}, \bibinfo
  {author} {\bibfnamefont {L.~W.}\ \bibnamefont {Packer}}, \bibinfo {author}
  {\bibfnamefont {M.~R.}\ \bibnamefont {Gilbert}}, \bibinfo {author}
  {\bibfnamefont {M.}~\bibnamefont {Klimenkov}}, \bibinfo {author}
  {\bibfnamefont {M.}~\bibnamefont {Rieth}}, \bibinfo {author} {\bibfnamefont
  {H.-C.}\ \bibnamefont {Schneider}}, \bibinfo {author} {\bibfnamefont {C.~D.}\
  \bibnamefont {Hardie}}, \bibinfo {author} {\bibfnamefont {S.~G.}\
  \bibnamefont {Roberts}},\ and\ \bibinfo {author} {\bibfnamefont {D.~E.~J.}\
  \bibnamefont {Armstrong}},\ }\bibfield  {title} {\bibinfo {title} {Effects of
  neutron irradiation on the brittle to ductile transition in single crystal
  tungsten},\ }\href@noop {} {\bibfield  {journal} {\bibinfo  {journal}
  {Journal of Nuclear Materials}\ }\textbf {\bibinfo {volume} {527}},\ \bibinfo
  {pages} {151799} (\bibinfo {year} {2019})}\BibitemShut {NoStop}%
\bibitem [{\citenamefont {Swinburne}\ and\ \citenamefont
  {Dudarev}(2018)}]{swinburne2018kink}%
  \BibitemOpen
  \bibfield  {author} {\bibinfo {author} {\bibfnamefont {T.~D.}\ \bibnamefont
  {Swinburne}}\ and\ \bibinfo {author} {\bibfnamefont {S.~L.}\ \bibnamefont
  {Dudarev}},\ }\bibfield  {title} {\bibinfo {title} {Kink-limited {Orowan}
  strengthening explains the brittle to ductile transition of irradiated and
  unirradiated bcc metals},\ }\href@noop {} {\bibfield  {journal} {\bibinfo
  {journal} {Physical Review Materials}\ }\textbf {\bibinfo {volume} {2}},\
  \bibinfo {pages} {073608} (\bibinfo {year} {2018})}\BibitemShut {NoStop}%
\bibitem [{\citenamefont {Hirth}\ and\ \citenamefont
  {Lothe}(1982)}]{hirth1982theory}%
  \BibitemOpen
  \bibfield  {author} {\bibinfo {author} {\bibfnamefont {J.~P.}\ \bibnamefont
  {Hirth}}\ and\ \bibinfo {author} {\bibfnamefont {J.}~\bibnamefont {Lothe}},\
  }\href@noop {} {\emph {\bibinfo {title} {Theory of Dislocations}}}\ (\bibinfo
   {publisher} {Wiley},\ \bibinfo {address} {New York},\ \bibinfo {year}
  {1982})\ Chap.~\bibinfo {chapter} {14}\BibitemShut {NoStop}%
\bibitem [{\citenamefont {Maeda}\ and\ \citenamefont
  {Yamashita}(1993)}]{maeda1993dislocation}%
  \BibitemOpen
  \bibfield  {author} {\bibinfo {author} {\bibfnamefont {K.}~\bibnamefont
  {Maeda}}\ and\ \bibinfo {author} {\bibfnamefont {Y.}~\bibnamefont
  {Yamashita}},\ }\bibfield  {title} {\bibinfo {title} {Dislocation motion in
  strained thin films. are kinks colliding with each other?},\ }\href@noop {}
  {\bibfield  {journal} {\bibinfo  {journal} {Physica Status Solidi (a)}\
  }\textbf {\bibinfo {volume} {138}},\ \bibinfo {pages} {523} (\bibinfo {year}
  {1993})}\BibitemShut {NoStop}%
\bibitem [{\citenamefont {Robertson}\ and\ \citenamefont
  {Gururaj}(2011)}]{Robertson2011}%
  \BibitemOpen
  \bibfield  {author} {\bibinfo {author} {\bibfnamefont {C.}~\bibnamefont
  {Robertson}}\ and\ \bibinfo {author} {\bibfnamefont {K.}~\bibnamefont
  {Gururaj}},\ }\bibfield  {title} {\bibinfo {title} {Plastic deformation of
  ferritic grains in presence of {ODS} particles and irradiation-induced defect
  clusters: A {3D} dislocation dynamics simulation study},\ }\href@noop {}
  {\bibfield  {journal} {\bibinfo  {journal} {Journal of Nuclear Materials}\
  }\textbf {\bibinfo {volume} {415}},\ \bibinfo {pages} {167–178} (\bibinfo
  {year} {2011})}\BibitemShut {NoStop}%
\bibitem [{\citenamefont {Cai}\ and\ \citenamefont
  {Bulatov}(2004)}]{cai2004mobility}%
  \BibitemOpen
  \bibfield  {author} {\bibinfo {author} {\bibfnamefont {W.}~\bibnamefont
  {Cai}}\ and\ \bibinfo {author} {\bibfnamefont {V.~V.}\ \bibnamefont
  {Bulatov}},\ }\bibfield  {title} {\bibinfo {title} {Mobility laws in
  dislocation dynamics simulations},\ }\href@noop {} {\bibfield  {journal}
  {\bibinfo  {journal} {Materials Science and Engineering: A}\ }\textbf
  {\bibinfo {volume} {387}},\ \bibinfo {pages} {277} (\bibinfo {year}
  {2004})}\BibitemShut {NoStop}%
\bibitem [{\citenamefont {Monnet}\ \emph {et~al.}(2013)\citenamefont {Monnet},
  \citenamefont {Vincent},\ and\ \citenamefont
  {Devincre}}]{monnet2013dislocation}%
  \BibitemOpen
  \bibfield  {author} {\bibinfo {author} {\bibfnamefont {G.}~\bibnamefont
  {Monnet}}, \bibinfo {author} {\bibfnamefont {L.}~\bibnamefont {Vincent}},\
  and\ \bibinfo {author} {\bibfnamefont {B.}~\bibnamefont {Devincre}},\
  }\bibfield  {title} {\bibinfo {title} {Dislocation-dynamics based crystal
  plasticity law for the low-and high-temperature deformation regimes of bcc
  crystal},\ }\href@noop {} {\bibfield  {journal} {\bibinfo  {journal} {Acta
  Materialia}\ }\textbf {\bibinfo {volume} {61}},\ \bibinfo {pages} {6178}
  (\bibinfo {year} {2013})}\BibitemShut {NoStop}%
\bibitem [{\citenamefont {Vattr{\'e}}\ \emph {et~al.}(2014)\citenamefont
  {Vattr{\'e}}, \citenamefont {Devincre}, \citenamefont {Feyel}, \citenamefont
  {Gatti}, \citenamefont {Groh}, \citenamefont {Jamond},\ and\ \citenamefont
  {Roos}}]{vattre2014modelling}%
  \BibitemOpen
  \bibfield  {author} {\bibinfo {author} {\bibfnamefont {A.}~\bibnamefont
  {Vattr{\'e}}}, \bibinfo {author} {\bibfnamefont {B.}~\bibnamefont
  {Devincre}}, \bibinfo {author} {\bibfnamefont {F.}~\bibnamefont {Feyel}},
  \bibinfo {author} {\bibfnamefont {R.}~\bibnamefont {Gatti}}, \bibinfo
  {author} {\bibfnamefont {S.}~\bibnamefont {Groh}}, \bibinfo {author}
  {\bibfnamefont {O.}~\bibnamefont {Jamond}},\ and\ \bibinfo {author}
  {\bibfnamefont {A.}~\bibnamefont {Roos}},\ }\bibfield  {title} {\bibinfo
  {title} {Modelling crystal plasticity by 3d dislocation dynamics and the
  finite element method: the discrete-continuous model revisited},\ }\href@noop
  {} {\bibfield  {journal} {\bibinfo  {journal} {Journal of the Mechanics and
  Physics of Solids}\ }\textbf {\bibinfo {volume} {63}},\ \bibinfo {pages}
  {491} (\bibinfo {year} {2014})}\BibitemShut {NoStop}%
\bibitem [{\citenamefont {Cereceda}\ \emph {et~al.}(2016)\citenamefont
  {Cereceda}, \citenamefont {Diehl}, \citenamefont {Roters}, \citenamefont
  {Raabe}, \citenamefont {Perlado},\ and\ \citenamefont
  {Marian}}]{cereceda2016unraveling}%
  \BibitemOpen
  \bibfield  {author} {\bibinfo {author} {\bibfnamefont {D.}~\bibnamefont
  {Cereceda}}, \bibinfo {author} {\bibfnamefont {M.}~\bibnamefont {Diehl}},
  \bibinfo {author} {\bibfnamefont {F.}~\bibnamefont {Roters}}, \bibinfo
  {author} {\bibfnamefont {D.}~\bibnamefont {Raabe}}, \bibinfo {author}
  {\bibfnamefont {J.~M.}\ \bibnamefont {Perlado}},\ and\ \bibinfo {author}
  {\bibfnamefont {J.}~\bibnamefont {Marian}},\ }\bibfield  {title} {\bibinfo
  {title} {Unraveling the temperature dependence of the yield strength in
  single-crystal tungsten using atomistically-informed crystal plasticity
  calculations},\ }\href@noop {} {\bibfield  {journal} {\bibinfo  {journal}
  {International Journal of Plasticity}\ }\textbf {\bibinfo {volume} {78}},\
  \bibinfo {pages} {242} (\bibinfo {year} {2016})}\BibitemShut {NoStop}%
\bibitem [{Note1()}]{Note1}%
  \BibitemOpen
  \bibinfo {note} {{Left and right kinks on a $\protect \frac {1}{2}\left
  \delimiter "426830A 111 \right \delimiter "526930B $ bcc screw dislocation
  have different formation energies.\cite {bulatov1997kink,
  ventelon2009atomistic} Using the average formation energy is not an
  approximation because kinks are only formed in pairs, provided periodic
  boundary conditions apply.}}\BibitemShut {Stop}%
\bibitem [{\citenamefont {Swinburne}\ and\ \citenamefont
  {Marinica}(2018)}]{swinburne2018unsupervised}%
  \BibitemOpen
  \bibfield  {author} {\bibinfo {author} {\bibfnamefont {T.~D.}\ \bibnamefont
  {Swinburne}}\ and\ \bibinfo {author} {\bibfnamefont {M.-C.}\ \bibnamefont
  {Marinica}},\ }\bibfield  {title} {\bibinfo {title} {Unsupervised calculation
  of free energy barriers in large crystalline systems},\ }\href@noop {}
  {\bibfield  {journal} {\bibinfo  {journal} {Physical Review Letters}\
  }\textbf {\bibinfo {volume} {120}},\ \bibinfo {pages} {135503} (\bibinfo
  {year} {2018})}\BibitemShut {NoStop}%
\bibitem [{\citenamefont {Chui}\ and\ \citenamefont
  {Weeks}(1976)}]{chui1976phase}%
  \BibitemOpen
  \bibfield  {author} {\bibinfo {author} {\bibfnamefont {S.~T.}\ \bibnamefont
  {Chui}}\ and\ \bibinfo {author} {\bibfnamefont {J.~D.}\ \bibnamefont
  {Weeks}},\ }\bibfield  {title} {\bibinfo {title} {Phase transition in the
  two-dimensional {Coulomb} gas, and the interfacial roughening transition},\
  }\href@noop {} {\bibfield  {journal} {\bibinfo  {journal} {Physical Review
  B}\ }\textbf {\bibinfo {volume} {14}},\ \bibinfo {pages} {4978} (\bibinfo
  {year} {1976})}\BibitemShut {NoStop}%
\bibitem [{\citenamefont {Beukers}(2007)}]{beukers2007gauss}%
  \BibitemOpen
  \bibfield  {author} {\bibinfo {author} {\bibfnamefont {F.}~\bibnamefont
  {Beukers}},\ }\bibfield  {title} {\bibinfo {title} {Gauss' hypergeometric
  function},\ }in\ \href@noop {} {\emph {\bibinfo {booktitle} {Arithmetic and
  geometry around hypergeometric functions}}},\ \bibinfo {editor} {edited by\
  \bibinfo {editor} {\bibfnamefont {R.-P.}\ \bibnamefont {Holzapfel}}, \bibinfo
  {editor} {\bibfnamefont {M.}~\bibnamefont {Yoshida}},\ and\ \bibinfo {editor}
  {\bibfnamefont {A.~M.}\ \bibnamefont {Uluda\u{g}}}}\ (\bibinfo  {publisher}
  {Birkh\"{a}user},\ \bibinfo {address} {Basel},\ \bibinfo {year} {2007})\ pp.\
  \bibinfo {pages} {23--42}\BibitemShut {NoStop}%
\bibitem [{\citenamefont {Seeger}(2004)}]{seeger2004}%
  \BibitemOpen
  \bibfield  {author} {\bibinfo {author} {\bibfnamefont {A.}~\bibnamefont
  {Seeger}},\ }\bibfield  {title} {\bibinfo {title} {Progress and problems in
  the understanding of the dislocation relaxation processes in metals},\
  }\href@noop {} {\bibfield  {journal} {\bibinfo  {journal} {Materials Science
  and Engineering A}\ }\textbf {\bibinfo {volume} {370}},\ \bibinfo {pages}
  {50–66} (\bibinfo {year} {2004})}\BibitemShut {NoStop}%
\bibitem [{\citenamefont {Po}\ \emph {et~al.}(2016)\citenamefont {Po},
  \citenamefont {Cui}, \citenamefont {Rivera}, \citenamefont {Cereceda},
  \citenamefont {Swinburne}, \citenamefont {Marian},\ and\ \citenamefont
  {Ghoniem}}]{po2016phenomenological}%
  \BibitemOpen
  \bibfield  {author} {\bibinfo {author} {\bibfnamefont {G.}~\bibnamefont
  {Po}}, \bibinfo {author} {\bibfnamefont {Y.}~\bibnamefont {Cui}}, \bibinfo
  {author} {\bibfnamefont {D.}~\bibnamefont {Rivera}}, \bibinfo {author}
  {\bibfnamefont {D.}~\bibnamefont {Cereceda}}, \bibinfo {author}
  {\bibfnamefont {T.~D.}\ \bibnamefont {Swinburne}}, \bibinfo {author}
  {\bibfnamefont {J.}~\bibnamefont {Marian}},\ and\ \bibinfo {author}
  {\bibfnamefont {N.}~\bibnamefont {Ghoniem}},\ }\bibfield  {title} {\bibinfo
  {title} {A phenomenological dislocation mobility law for bcc metals},\
  }\href@noop {} {\bibfield  {journal} {\bibinfo  {journal} {Acta Materialia}\
  }\textbf {\bibinfo {volume} {119}},\ \bibinfo {pages} {123} (\bibinfo {year}
  {2016})}\BibitemShut {NoStop}%
\bibitem [{\citenamefont {Ares}\ \emph {et~al.}(2003)\citenamefont {Ares},
  \citenamefont {Cuesta}, \citenamefont {S{\'a}nchez},\ and\ \citenamefont
  {Toral}}]{ares2003apparent}%
  \BibitemOpen
  \bibfield  {author} {\bibinfo {author} {\bibfnamefont {S.}~\bibnamefont
  {Ares}}, \bibinfo {author} {\bibfnamefont {J.}~\bibnamefont {Cuesta}},
  \bibinfo {author} {\bibfnamefont {A.}~\bibnamefont {S{\'a}nchez}},\ and\
  \bibinfo {author} {\bibfnamefont {R.}~\bibnamefont {Toral}},\ }\bibfield
  {title} {\bibinfo {title} {Apparent phase transitions in finite
  one-dimensional sine-{Gordo}n lattices},\ }\href@noop {} {\bibfield
  {journal} {\bibinfo  {journal} {Physical Review E}\ }\textbf {\bibinfo
  {volume} {67}},\ \bibinfo {pages} {046108} (\bibinfo {year}
  {2003})}\BibitemShut {NoStop}%
\bibitem [{\citenamefont {Baskaran}\ and\ \citenamefont
  {Gupte}(1984)}]{baskaran1984equivalence}%
  \BibitemOpen
  \bibfield  {author} {\bibinfo {author} {\bibfnamefont {G.}~\bibnamefont
  {Baskaran}}\ and\ \bibinfo {author} {\bibfnamefont {N.}~\bibnamefont
  {Gupte}},\ }\bibfield  {title} {\bibinfo {title} {Equivalence between the
  discrete {Gaussian} model and a generalized sine-{Gordon} theory on a
  lattice},\ }\href@noop {} {\bibfield  {journal} {\bibinfo  {journal}
  {Physical Review B}\ }\textbf {\bibinfo {volume} {30}},\ \bibinfo {pages}
  {432} (\bibinfo {year} {1984})}\BibitemShut {NoStop}%
\bibitem [{\citenamefont {Basharin}\ \emph {et~al.}(2004)\citenamefont
  {Basharin}, \citenamefont {Langville},\ and\ \citenamefont
  {Naumov}}]{Markov}%
  \BibitemOpen
  \bibfield  {author} {\bibinfo {author} {\bibfnamefont {G.~P.}\ \bibnamefont
  {Basharin}}, \bibinfo {author} {\bibfnamefont {A.~N.}\ \bibnamefont
  {Langville}},\ and\ \bibinfo {author} {\bibfnamefont {V.~A.}\ \bibnamefont
  {Naumov}},\ }\bibfield  {title} {\bibinfo {title} {The life and work of {A.
  A. Markov}},\ }\href@noop {} {\bibfield  {journal} {\bibinfo  {journal}
  {Linear Algebra and its Applications}\ }\textbf {\bibinfo {volume} {386}},\
  \bibinfo {pages} {3–26} (\bibinfo {year} {2004})}\BibitemShut {NoStop}%
\bibitem [{Note2()}]{Note2}%
  \BibitemOpen
  \bibinfo {note} {Determining the group size $N_{r,A}$ is equivalent to the
  lattice enumeration problem\cite {bona2015handbook} of finding the number of
  grand Motzkin paths from $(0,0)$ to $(n,0)$ weighted by up (or down) steps
  and area. We refer to some examples here \cite {owczarek2009exact,
  owczarek2010exact}. This is also required to determine the total number of
  reactions of each $(r,A)$ group into the connected groups.}\BibitemShut
  {Stop}%
\bibitem [{\citenamefont {Voter}(2007)}]{voter2007introduction}%
  \BibitemOpen
  \bibfield  {author} {\bibinfo {author} {\bibfnamefont {A.~F.}\ \bibnamefont
  {Voter}},\ }\bibfield  {title} {\bibinfo {title} {Introduction to the kinetic
  {Monte} {Carlo} method},\ }in\ \href@noop {} {\emph {\bibinfo {booktitle}
  {Radiation Effects in Solids}}},\ \bibinfo {editor} {edited by\ \bibinfo
  {editor} {\bibfnamefont {K.~E.}\ \bibnamefont {Sickafus}}, \bibinfo {editor}
  {\bibfnamefont {E.~A.}\ \bibnamefont {Kotomin}},\ and\ \bibinfo {editor}
  {\bibfnamefont {B.~P.}\ \bibnamefont {Uberuaga}}}\ (\bibinfo  {publisher}
  {Springer},\ \bibinfo {address} {Dordrecht},\ \bibinfo {year} {2007})\
  Chap.~\bibinfo {chapter} {1}, pp.\ \bibinfo {pages} {1--23}\BibitemShut
  {NoStop}%
\bibitem [{\citenamefont {Gordon}\ \emph {et~al.}(2011)\citenamefont {Gordon},
  \citenamefont {Neeraj},\ and\ \citenamefont {Mendelev}}]{gordon2011screw}%
  \BibitemOpen
  \bibfield  {author} {\bibinfo {author} {\bibfnamefont {P.~A.}\ \bibnamefont
  {Gordon}}, \bibinfo {author} {\bibfnamefont {T.}~\bibnamefont {Neeraj}},\
  and\ \bibinfo {author} {\bibfnamefont {M.~I.}\ \bibnamefont {Mendelev}},\
  }\bibfield  {title} {\bibinfo {title} {Screw dislocation mobility in bcc
  metals: a refined potential description for $\alpha$-{Fe}},\ }\href@noop {}
  {\bibfield  {journal} {\bibinfo  {journal} {Philosophical Magazine}\ }\textbf
  {\bibinfo {volume} {91}},\ \bibinfo {pages} {3931} (\bibinfo {year}
  {2011})}\BibitemShut {NoStop}%
\bibitem [{\citenamefont {Gilbert}\ \emph {et~al.}(2013)\citenamefont
  {Gilbert}, \citenamefont {Schuck}, \citenamefont {Sadigh},\ and\
  \citenamefont {Marian}}]{gilbert2013free}%
  \BibitemOpen
  \bibfield  {author} {\bibinfo {author} {\bibfnamefont {M.~R.}\ \bibnamefont
  {Gilbert}}, \bibinfo {author} {\bibfnamefont {P.}~\bibnamefont {Schuck}},
  \bibinfo {author} {\bibfnamefont {B.}~\bibnamefont {Sadigh}},\ and\ \bibinfo
  {author} {\bibfnamefont {J.}~\bibnamefont {Marian}},\ }\bibfield  {title}
  {\bibinfo {title} {Free energy generalization of the peierls potential in
  iron},\ }\href@noop {} {\bibfield  {journal} {\bibinfo  {journal} {Physical
  Review Letters}\ }\textbf {\bibinfo {volume} {111}},\ \bibinfo {pages}
  {095502} (\bibinfo {year} {2013})}\BibitemShut {NoStop}%
\bibitem [{\citenamefont {Proville}\ \emph {et~al.}(2012)\citenamefont
  {Proville}, \citenamefont {Rodney},\ and\ \citenamefont
  {Marinica}}]{proville2012quantum}%
  \BibitemOpen
  \bibfield  {author} {\bibinfo {author} {\bibfnamefont {L.}~\bibnamefont
  {Proville}}, \bibinfo {author} {\bibfnamefont {D.}~\bibnamefont {Rodney}},\
  and\ \bibinfo {author} {\bibfnamefont {M.-C.}\ \bibnamefont {Marinica}},\
  }\bibfield  {title} {\bibinfo {title} {Quantum effect on thermally activated
  glide of dislocations},\ }\href@noop {} {\bibfield  {journal} {\bibinfo
  {journal} {Nature materials}\ }\textbf {\bibinfo {volume} {11}},\ \bibinfo
  {pages} {845} (\bibinfo {year} {2012})}\BibitemShut {NoStop}%
\bibitem [{\citenamefont {Stukowski}\ \emph {et~al.}(2015)\citenamefont
  {Stukowski}, \citenamefont {Cereceda}, \citenamefont {Swinburne},\ and\
  \citenamefont {Marian}}]{stukowski2015thermally}%
  \BibitemOpen
  \bibfield  {author} {\bibinfo {author} {\bibfnamefont {A.}~\bibnamefont
  {Stukowski}}, \bibinfo {author} {\bibfnamefont {D.}~\bibnamefont {Cereceda}},
  \bibinfo {author} {\bibfnamefont {T.~D.}\ \bibnamefont {Swinburne}},\ and\
  \bibinfo {author} {\bibfnamefont {J.}~\bibnamefont {Marian}},\ }\bibfield
  {title} {\bibinfo {title} {Thermally-activated non-{Schmid} glide of screw
  dislocations in {W} using atomistically-informed kinetic {Monte} {Carlo}
  simulations},\ }\href@noop {} {\bibfield  {journal} {\bibinfo  {journal}
  {International Journal of Plasticity}\ }\textbf {\bibinfo {volume} {65}},\
  \bibinfo {pages} {108} (\bibinfo {year} {2015})}\BibitemShut {NoStop}%
\bibitem [{\citenamefont {Marinica}\ \emph {et~al.}(2013)\citenamefont
  {Marinica}, \citenamefont {Ventelon}, \citenamefont {Gilbert}, \citenamefont
  {Proville}, \citenamefont {Dudarev}, \citenamefont {Marian}, \citenamefont
  {Bencteux},\ and\ \citenamefont {Willaime}}]{marinica2013interatomic}%
  \BibitemOpen
  \bibfield  {author} {\bibinfo {author} {\bibfnamefont {M.-C.}\ \bibnamefont
  {Marinica}}, \bibinfo {author} {\bibfnamefont {L.}~\bibnamefont {Ventelon}},
  \bibinfo {author} {\bibfnamefont {M.~R.}\ \bibnamefont {Gilbert}}, \bibinfo
  {author} {\bibfnamefont {L.}~\bibnamefont {Proville}}, \bibinfo {author}
  {\bibfnamefont {S.~L.}\ \bibnamefont {Dudarev}}, \bibinfo {author}
  {\bibfnamefont {J.}~\bibnamefont {Marian}}, \bibinfo {author} {\bibfnamefont
  {G.}~\bibnamefont {Bencteux}},\ and\ \bibinfo {author} {\bibfnamefont
  {F.}~\bibnamefont {Willaime}},\ }\bibfield  {title} {\bibinfo {title}
  {Interatomic potentials for modelling radiation defects and dislocations in
  tungsten},\ }\href@noop {} {\bibfield  {journal} {\bibinfo  {journal}
  {Journal of Physics: Condensed Matter}\ }\textbf {\bibinfo {volume} {25}},\
  \bibinfo {pages} {395502} (\bibinfo {year} {2013})}\BibitemShut {NoStop}%
\bibitem [{\citenamefont {Suzuki}\ \emph {et~al.}(1999)\citenamefont {Suzuki},
  \citenamefont {Kamimura},\ and\ \citenamefont
  {Kirchner}}]{suzuki1999plastic}%
  \BibitemOpen
  \bibfield  {author} {\bibinfo {author} {\bibfnamefont {T.}~\bibnamefont
  {Suzuki}}, \bibinfo {author} {\bibfnamefont {Y.}~\bibnamefont {Kamimura}},\
  and\ \bibinfo {author} {\bibfnamefont {H.~O.~K.}\ \bibnamefont {Kirchner}},\
  }\bibfield  {title} {\bibinfo {title} {Plastic homology of bcc metals},\
  }\href@noop {} {\bibfield  {journal} {\bibinfo  {journal} {Philosophical
  Magazine A}\ }\textbf {\bibinfo {volume} {79}},\ \bibinfo {pages} {1629}
  (\bibinfo {year} {1999})}\BibitemShut {NoStop}%
\bibitem [{\citenamefont {Kamimura}\ \emph {et~al.}(2013)\citenamefont
  {Kamimura}, \citenamefont {Edagawa},\ and\ \citenamefont
  {Takeuchi}}]{kamimura2013experimental}%
  \BibitemOpen
  \bibfield  {author} {\bibinfo {author} {\bibfnamefont {Y.}~\bibnamefont
  {Kamimura}}, \bibinfo {author} {\bibfnamefont {K.}~\bibnamefont {Edagawa}},\
  and\ \bibinfo {author} {\bibfnamefont {S.}~\bibnamefont {Takeuchi}},\
  }\bibfield  {title} {\bibinfo {title} {Experimental evaluation of the
  {Peierls} stresses in a variety of crystals and their relation to the crystal
  structure},\ }\href@noop {} {\bibfield  {journal} {\bibinfo  {journal} {Acta
  materialia}\ }\textbf {\bibinfo {volume} {61}},\ \bibinfo {pages} {294}
  (\bibinfo {year} {2013})}\BibitemShut {NoStop}%
\bibitem [{\citenamefont {Seeger}\ and\ \citenamefont
  {Holzwarth}(2006)}]{seeger2006slip}%
  \BibitemOpen
  \bibfield  {author} {\bibinfo {author} {\bibfnamefont {A.}~\bibnamefont
  {Seeger}}\ and\ \bibinfo {author} {\bibfnamefont {U.}~\bibnamefont
  {Holzwarth}},\ }\bibfield  {title} {\bibinfo {title} {Slip planes and kink
  properties of screw dislocations in high-purity niobium},\ }\href@noop {}
  {\bibfield  {journal} {\bibinfo  {journal} {Philosophical Magazine}\ }\textbf
  {\bibinfo {volume} {86}},\ \bibinfo {pages} {3861} (\bibinfo {year}
  {2006})}\BibitemShut {NoStop}%
\bibitem [{\citenamefont {Rodney}\ and\ \citenamefont
  {Proville}(2009)}]{rodney2009stress}%
  \BibitemOpen
  \bibfield  {author} {\bibinfo {author} {\bibfnamefont {D.}~\bibnamefont
  {Rodney}}\ and\ \bibinfo {author} {\bibfnamefont {L.}~\bibnamefont
  {Proville}},\ }\bibfield  {title} {\bibinfo {title} {Stress-dependent
  {Peierls} potential: Influence on kink-pair activation},\ }\href@noop {}
  {\bibfield  {journal} {\bibinfo  {journal} {Physical Review B}\ }\textbf
  {\bibinfo {volume} {79}},\ \bibinfo {pages} {094108} (\bibinfo {year}
  {2009})}\BibitemShut {NoStop}%
\bibitem [{\citenamefont {Proville}\ \emph {et~al.}(2013)\citenamefont
  {Proville}, \citenamefont {Ventelon},\ and\ \citenamefont
  {Rodney}}]{proville2013prediction}%
  \BibitemOpen
  \bibfield  {author} {\bibinfo {author} {\bibfnamefont {L.}~\bibnamefont
  {Proville}}, \bibinfo {author} {\bibfnamefont {L.}~\bibnamefont {Ventelon}},\
  and\ \bibinfo {author} {\bibfnamefont {D.}~\bibnamefont {Rodney}},\
  }\bibfield  {title} {\bibinfo {title} {Prediction of the kink-pair formation
  enthalpy on screw dislocations in $\alpha$-iron by a line tension model
  parametrized on empirical potentials and first-principles calculations},\
  }\href@noop {} {\bibfield  {journal} {\bibinfo  {journal} {Physical Review
  B}\ }\textbf {\bibinfo {volume} {87}},\ \bibinfo {pages} {144106} (\bibinfo
  {year} {2013})}\BibitemShut {NoStop}%
\bibitem [{\citenamefont {Maresca}\ \emph {et~al.}(2018)\citenamefont
  {Maresca}, \citenamefont {Dragoni}, \citenamefont {Cs{\'a}nyi}, \citenamefont
  {Marzari},\ and\ \citenamefont {Curtin}}]{maresca2018screw}%
  \BibitemOpen
  \bibfield  {author} {\bibinfo {author} {\bibfnamefont {F.}~\bibnamefont
  {Maresca}}, \bibinfo {author} {\bibfnamefont {D.}~\bibnamefont {Dragoni}},
  \bibinfo {author} {\bibfnamefont {G.}~\bibnamefont {Cs{\'a}nyi}}, \bibinfo
  {author} {\bibfnamefont {N.}~\bibnamefont {Marzari}},\ and\ \bibinfo {author}
  {\bibfnamefont {W.~A.}\ \bibnamefont {Curtin}},\ }\bibfield  {title}
  {\bibinfo {title} {Screw dislocation structure and mobility in body centered
  cubic {Fe} predicted by a {Gaussian} {Approximation} {Potential}},\
  }\href@noop {} {\bibfield  {journal} {\bibinfo  {journal} {npj Computational
  Materials}\ }\textbf {\bibinfo {volume} {4}},\ \bibinfo {pages} {1} (\bibinfo
  {year} {2018})}\BibitemShut {NoStop}%
\bibitem [{\citenamefont {Gilbert}\ \emph {et~al.}(2011)\citenamefont
  {Gilbert}, \citenamefont {Queyreau},\ and\ \citenamefont
  {Marian}}]{gilbert2011stress}%
  \BibitemOpen
  \bibfield  {author} {\bibinfo {author} {\bibfnamefont {M.~R.}\ \bibnamefont
  {Gilbert}}, \bibinfo {author} {\bibfnamefont {S.}~\bibnamefont {Queyreau}},\
  and\ \bibinfo {author} {\bibfnamefont {J.}~\bibnamefont {Marian}},\
  }\bibfield  {title} {\bibinfo {title} {Stress and temperature dependence of
  screw dislocation mobility in $\alpha$-{Fe} by molecular dynamics},\
  }\href@noop {} {\bibfield  {journal} {\bibinfo  {journal} {Physical Review
  B}\ }\textbf {\bibinfo {volume} {84}},\ \bibinfo {pages} {174103} (\bibinfo
  {year} {2011})}\BibitemShut {NoStop}%
\bibitem [{\citenamefont {Bulatov}\ \emph {et~al.}(1997)\citenamefont
  {Bulatov}, \citenamefont {Justo}, \citenamefont {Cai},\ and\ \citenamefont
  {Yip}}]{bulatov1997kink}%
  \BibitemOpen
  \bibfield  {author} {\bibinfo {author} {\bibfnamefont {V.~V.}\ \bibnamefont
  {Bulatov}}, \bibinfo {author} {\bibfnamefont {J.~F.}\ \bibnamefont {Justo}},
  \bibinfo {author} {\bibfnamefont {W.}~\bibnamefont {Cai}},\ and\ \bibinfo
  {author} {\bibfnamefont {S.}~\bibnamefont {Yip}},\ }\bibfield  {title}
  {\bibinfo {title} {Kink asymmetry and multiplicity in dislocation cores},\
  }\href@noop {} {\bibfield  {journal} {\bibinfo  {journal} {Physical Review
  Letters}\ }\textbf {\bibinfo {volume} {79}},\ \bibinfo {pages} {5042}
  (\bibinfo {year} {1997})}\BibitemShut {NoStop}%
\bibitem [{\citenamefont {Ventelon}\ \emph {et~al.}(2009)\citenamefont
  {Ventelon}, \citenamefont {Willaime},\ and\ \citenamefont
  {Leyronnas}}]{ventelon2009atomistic}%
  \BibitemOpen
  \bibfield  {author} {\bibinfo {author} {\bibfnamefont {L.}~\bibnamefont
  {Ventelon}}, \bibinfo {author} {\bibfnamefont {F.}~\bibnamefont {Willaime}},\
  and\ \bibinfo {author} {\bibfnamefont {P.}~\bibnamefont {Leyronnas}},\
  }\bibfield  {title} {\bibinfo {title} {Atomistic simulation of single kinks
  of screw dislocations in $\alpha$-{Fe}},\ }\href@noop {} {\bibfield
  {journal} {\bibinfo  {journal} {Journal of Nuclear Materials}\ }\textbf
  {\bibinfo {volume} {386}},\ \bibinfo {pages} {26} (\bibinfo {year}
  {2009})}\BibitemShut {NoStop}%
\bibitem [{\citenamefont {Krattenthaler}(2015)}]{bona2015handbook}%
  \BibitemOpen
  \bibfield  {author} {\bibinfo {author} {\bibfnamefont {C.}~\bibnamefont
  {Krattenthaler}},\ }\bibfield  {title} {\bibinfo {title} {Handbook of
  enumerative combinatorics},\ }in\ \href@noop {} {\emph {\bibinfo {booktitle}
  {Handbook of Enumerative Combinatorics}}},\ \bibinfo {editor} {edited by\
  \bibinfo {editor} {\bibfnamefont {M.}~\bibnamefont {B{\'o}na}}}\ (\bibinfo
  {publisher} {CRC Press},\ \bibinfo {address} {New York},\ \bibinfo {year}
  {2015})\ Chap.~\bibinfo {chapter} {10}, pp.\ \bibinfo {pages}
  {589--670}\BibitemShut {NoStop}%
\bibitem [{\citenamefont {Owczarek}\ and\ \citenamefont
  {Prellberg}(2009)}]{owczarek2009exact}%
  \BibitemOpen
  \bibfield  {author} {\bibinfo {author} {\bibfnamefont {A.~L.}\ \bibnamefont
  {Owczarek}}\ and\ \bibinfo {author} {\bibfnamefont {T.}~\bibnamefont
  {Prellberg}},\ }\bibfield  {title} {\bibinfo {title} {Exact solution of the
  discrete (1+1)-dimensional {RSOS} model with field and surface
  interactions},\ }\href@noop {} {\bibfield  {journal} {\bibinfo  {journal}
  {Journal of Physics {A}: Mathematical and Theoretical}\ }\textbf {\bibinfo
  {volume} {42}},\ \bibinfo {pages} {495003} (\bibinfo {year}
  {2009})}\BibitemShut {NoStop}%
\bibitem [{\citenamefont {Owczarek}\ and\ \citenamefont
  {Prellberg}(2010)}]{owczarek2010exact}%
  \BibitemOpen
  \bibfield  {author} {\bibinfo {author} {\bibfnamefont {A.~L.}\ \bibnamefont
  {Owczarek}}\ and\ \bibinfo {author} {\bibfnamefont {T.}~\bibnamefont
  {Prellberg}},\ }\bibfield  {title} {\bibinfo {title} {Exact solution of the
  discrete (1+1)-dimensional {RSOS} model in a slit with field and wall
  interactions},\ }\href@noop {} {\bibfield  {journal} {\bibinfo  {journal}
  {Journal of Physics A: Mathematical and Theoretical}\ }\textbf {\bibinfo
  {volume} {43}},\ \bibinfo {pages} {375004} (\bibinfo {year}
  {2010})}\BibitemShut {NoStop}%
\end{thebibliography}%

\end{document}